\documentclass[apj]{emulateapj}

\usepackage{epsfig}
\usepackage{xspace}
\usepackage{color}

\newcommand{\rmn}{\mathrm}
\newcommand{\xvec}{{\bf{x}}}
\newcommand{\LAF}{\mathrm{Ly}\alpha}
\newcommand{\kb}{k_{\mathrm{o}}}
\newcommand{\bo}{b_{\mathrm{o}}}
\newcommand{\al}{\alpha}
\newcommand{\zre}{z_{\mathrm{RE}}({\bf{x}})}
\newcommand{\xe}{x_{\mathrm{e}}}
\newcommand{\zbar}{\bar{z}}
\newcommand{\rmz}{r_{\mathrm{mz}}}
\newcommand{\bmz}{b_{\mathrm{mz}} (k)}
\newcommand{\Delz}{\Delta_{\mathrm{z}}}
\newcommand{\lambe}{\lambda_{\mathrm{e}}}
\newcommand{\Mpch}{{\rm Mpc}/h}
\newcommand{\Msunh}{M_\odot/h}
\newcommand{\taue}{\tau_{\mathrm{e}}}
\newcommand{\te}{\tau}

\slugcomment{Submitted to ApJ}

\shorttitle{A Parametric model for Cosmic Reionization}
\shortauthors{Battaglia et al.}

\begin{document}

\title{Reionization on Large-Scales I: A Parametric Model Constructed from Radiation-Hydrodynamic Simulations}

\author{N. Battaglia\altaffilmark{1}, H. Trac\altaffilmark{1}, R. Cen\altaffilmark{2}, A. Loeb\altaffilmark{3}}

\altaffiltext{1}{McWilliams Center for Cosmology, Wean Hall, Carnegie Mellon University, 5000 Forbes Ave., Pittsburgh PA 15213, USA}
\altaffiltext{2}{Department of Astrophysical Sciences, Princeton University, Princeton, NJ 08544}
\altaffiltext{3}{Harvard-Smithsonian Center for Astrophysics, Cambridge, MA 02138}

\begin{abstract}

We present a new method for modeling inhomogeneous cosmic reionization on large scales. Utilizing high-resolution radiation-hydrodynamic simulations with 2048$^3$ dark matter particles, 2048$^3$ gas cells, and 17 billion adaptive rays in a L = 100 Mpc/$h$ box, we show that
the density and reionization-redshift fields are highly correlated on large scales ($\gtrsim 1\ $Mpc$/h$). This correlation can be statistically represented by a scale-dependent linear bias. We construct a parametric function for the bias, which is then used to filter any large-scale density field to derive the corresponding spatially varying reionization-redshift field. The parametric model has three free parameters which can be reduced to one free parameter when we fit the two bias parameters to simulations results. We can differentiate degenerate combinations of the bias parameters by combining results for the global ionization histories and correlation length between ionized regions.
Unlike previous semi-analytic models, the evolution of the reionization-redshift field in our model is directly compared cell by cell against simulations and preforms well in all tests. Our model maps the high-resolution, intermediate-volume radiation-hydrodynamic simulations onto lower-resolution, larger-volume N-body simulations ($\gtrsim 2$ Gpc$/h$) in order to make mock observations and theoretical predictions. 

\end{abstract}


\section{Introduction}

When the first stars and galaxies turned on and began ionizing the surrounding cold and neutral hydrogen of the intergalatic medium (IGM), they started the phase transition of the Universe known as the Epoch of Reionization \citep[EoR;][]{Loeb2013}. This inhomogeneous process of reionization leaves two, among others, possible observable sources, the neutral hydrogen atoms and the ionized electrons. It is necessary that precise theoretical models of EoR on Gpc scales are used to interpret the information from the observable imprints left by these sources in order to gain insight into and understanding of the first stars and the initial stages of galaxy formation \citep[e.g.][and references therein]{Furl2006,Mora2010}. 

Neutral hydrogen atoms are observed in both absorption and emission. Current constraints from absorption measurements come from observations zero transmission of rest-frame $\LAF$ flux at $z\gtrsim6$ in spectra of high redshift quasars, which suggest that EoR is completed by $z\sim6$ \citep{Fan2006a}, although it is possible for these constraints to be consistent with reionization completing at a higher redshift \citep[e.g.][]{Oh2005,Lidz2006}. The neutral hydrogen emission is observable through the redshifted 21cm signal that originates from its hyperfine transition \citep[e.g.][]{Scot1990,Shav1999,Zald2004}. There are several experiments currently searching for the 21cm signal at $z>6$, such as, the Murchison Wide Field Array \citep[MWA\footnote{www.mwatelescope.org};][]{Bow2005}, the Giant Meterwave Telescope \citep[GMRT\footnote{gmrt.ncra.tifr.res.in};][]{Pen2009}, the Low Frequency Array \citep[LOFAR\footnote{www.lofar.org};][]{Hark2010}, and The Precision Array for Probing the Epoch of Reionization \citep[PAPER\footnote{eor.berkeley.edu};][]{Pars2010}. Using the 21cm signal, the experiment EDGES\footnote{www.haystack.mit.edu/ast/arrays/Edges} reported a lower limit to duration of reionization, $\Delz > 0.06$ \citep{Bow2012}. Future 21 cm experiments like the proposed MeeRKAT\footnote{www.ska.ac.za/meerkat}\citep{Boot2009} and the Square Kilometer Array \citep[SKA\footnote{www.skatelescope.org};][]{Mell2012} both have the potential to measure this signal across several frequencies and provide tomographic information on EoR. 

\begin{figure*}
  \resizebox{0.5\hsize}{!}{\includegraphics{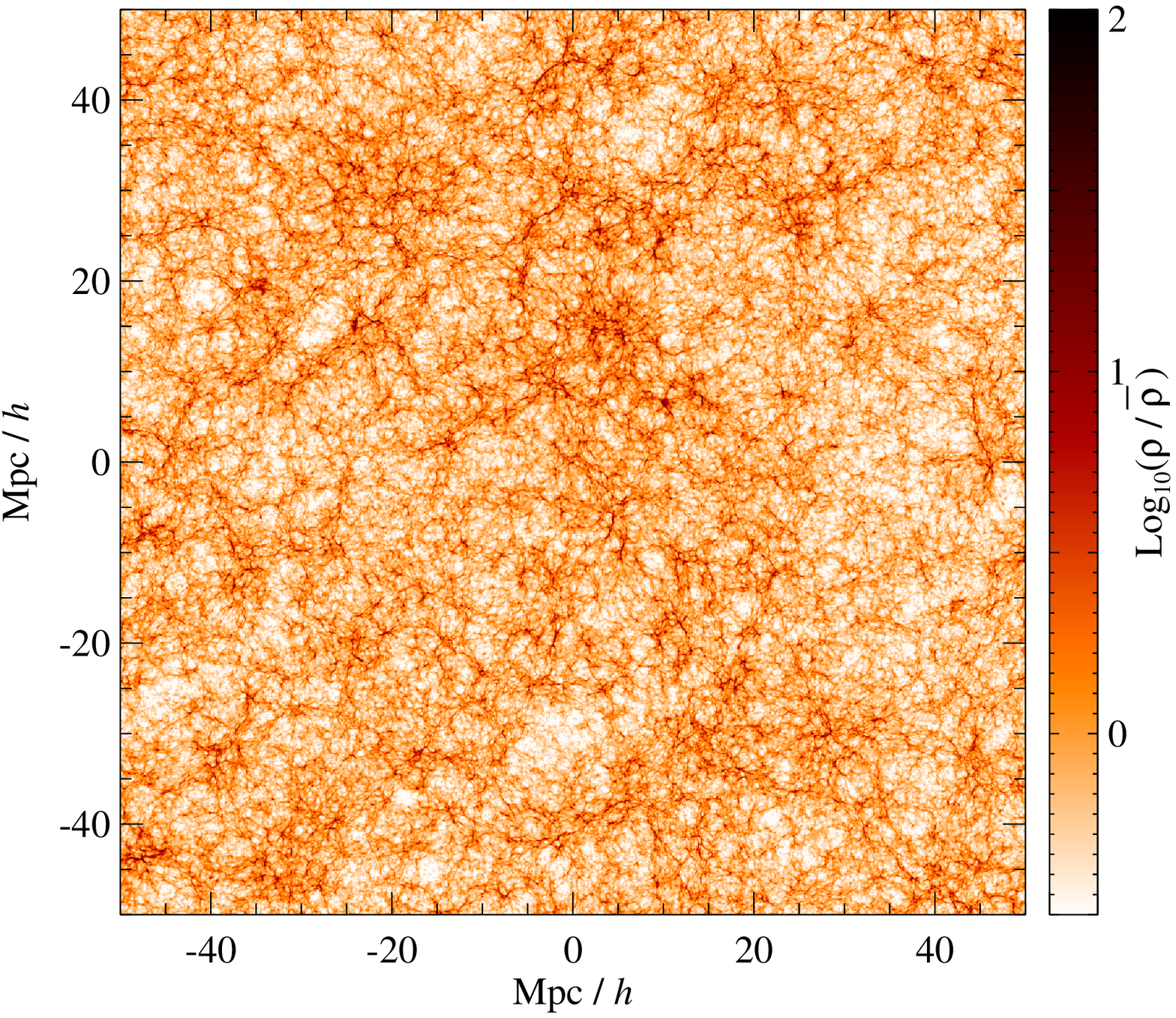}}%
  \resizebox{0.5\hsize}{!}{\includegraphics{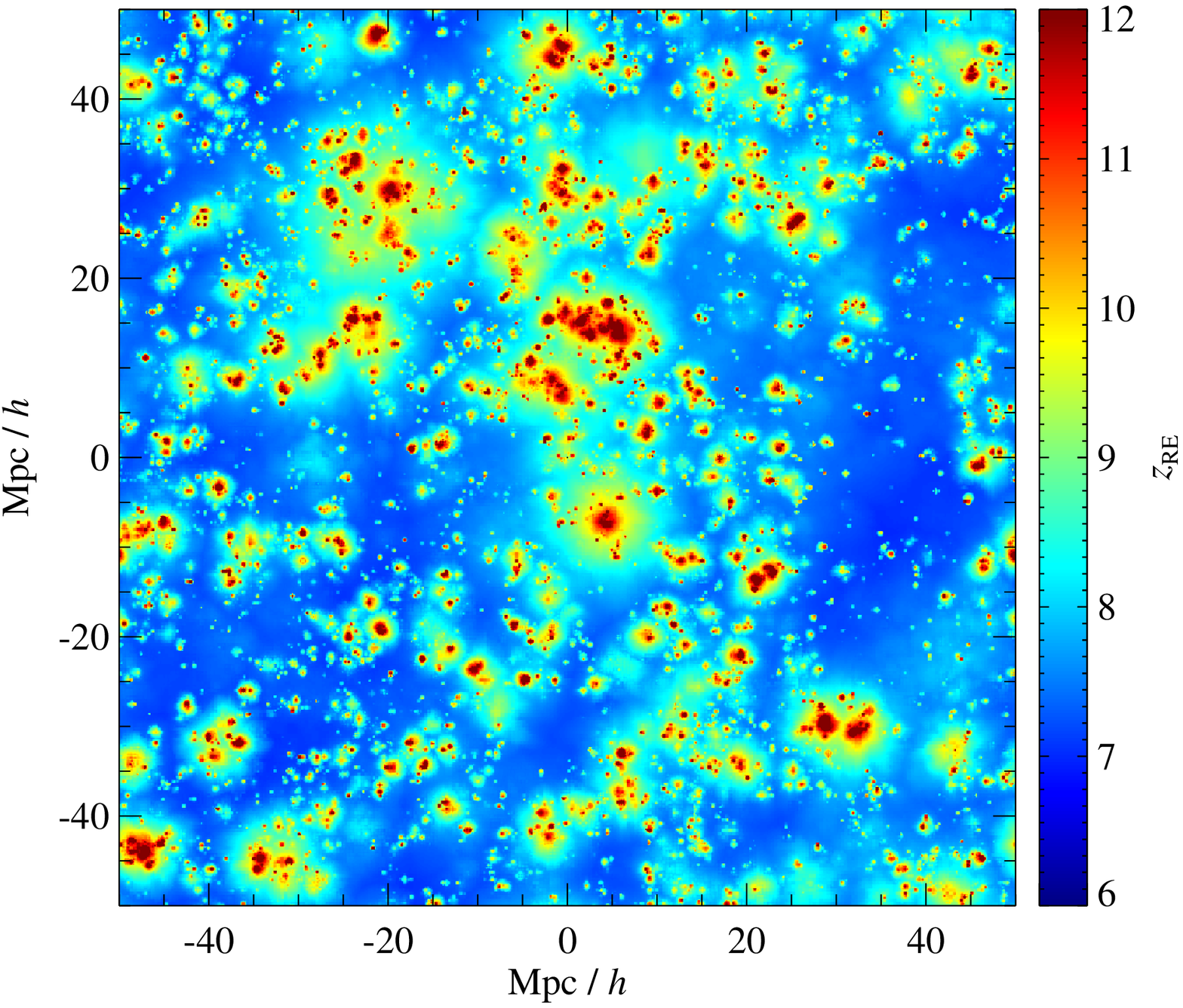}}\\ 
  \caption{Slices from our high resolution RadHydro simulation for a model of reionization that occurs ''late" with a midpoint of $z=8$ and is finished by $z \approx 6.9$. The dimensions are 100 Mpc/$h$ x 100 Mpc/$h$ with a thickness of $\sim$100 kpc/$h$ comoving. Left: The density field, $\rho(\xvec)/\bar{\rho}$. Right: The reionization-redshift field, $\zre$. Large-scale, overdense regions near sources are generally ionized earlier than large-scale, underdense regions far from sources.}
\label{fig:rho_z_sim}
\end{figure*}

The free electrons are observed through their scattering of cosmic microwave background (CMB) photons. This scattering can be seen on large scales in polarization CMB or on small scales in the CMB secondary anisotropies, such as kinetic Sunyaev Zel'dovich (kSZ) signal from the EoR \citep{Gruz1998,Knox1998,Vala2001,Sant2003,Zahn2005,McQn2005,iliv2007,Mess2011}. Polarization measurements of the CMB place a constraint on the optical depth to EoR \citep{WMAP7pars}. Assuming a step function or a hyperbolic tangent function for the ionization history, the 7-year WMAP data implies that the reionization-redshift is $10.5\pm1.2$ (68\% CL). Constraints have also come from multi-frequency high resolution CMB experiments that measure the power spectrum of CMB secondary anisotropies to great precision, where contributions to kSZ power from EOR are the largest. The South Pole Telescope \citep[SPT\footnote{pole.uchicago.edu};][]{Zahn2012} placed a model dependent upper limit on the duration of reionization from their multifrequency measurements of the high $\ell$ power spectrum, and future results from the Atacama Cosmology Telescope (ACT\footnote{www.princeton.edu/act}) are expected to place similar constraints. The next generation high resolution CMB experiments ACT with polarization (ACT-pol) and South Pole Telescope with polarization (SPT-pol) will precisely measurement the secondary anisotropies of the CMB in both temperature and polarization, which will provide tighter constraints on EoR.

For the EoR experiments listed above and future ones, the amount of understanding gained on these first ionizing sources and the initial stages of galaxy evolution will depend upon the accuracy of the theoretical models for EoR. The main challenge in EoR theory is providing an accurate model of the IGM, the sources and the sinks of ionizing photons, while having the a large enough volume $> 1$(Gpc/$h)^3$ to statistically sample the HI regions and construct mock observations on the angular scales required by the current and future EoR experiments.

There are two standard approaches to model EoR, radiative transfer simulations with various implementations for hydrodynamics and gas physics \citep[e.g.][]{Gned2001,Ciar2001,Mase2003,Alva2006,Mell2006,iliv2006,Trac2007,McQn2007,Trac2008,Aub2008,Alta2008,Crof2008,Fin2009,Pet2009} and semi-analytic models \citep[e.g.][]{Furl2004, Zahn2005, Zahn2007, Mess2007,Geil2008,Alva2009,Thom2009,Chod2009,Sant2010,Mess2011}.  In these semi-analytic models a region is fully ionized if the simple relation, $\zeta F_{\rmn{coll}} \ge 1$ is satisfied. Here $\zeta$ is an efficiency parameter and $F_{\rmn{coll}} $ is the collapse fraction, which is calculated via the excursion set formalism \citep{Bond1991}, or applied to three dimensional realization of a density field \citep[e.g.][]{Zahn2005}. Semi-analytic models capture the generic properties of EoR, but in order to capture the complex non-linear, and non-Gaussian nature of EoR radiative transfer simulations are required.

The advantages of the current full hydrodynamic, high resolution simulations with radiative transfer (implemented either in post processing or during the simulation) is that they probe the relevant scales to resolve sources of ionizing photons and their sinks, then trace these photons through an inhomogeneous IGM \citep{Trac2011}. However, full hydrodynamic simulations with radiative transfer on large enough scales to capture a representative sample of ionizing sources and with enough small scale resolution to also capture all the physics of reionization are currently not possible due to the overwhelming computational demands of such calculations. Thus, all of the simulations to date have been restricted to smaller box-sizes.
Recent work by \citet{Zahn2011} ran several convergence tests between these two types of EoR models. For all the models in their study, they found that the results from the models are within tens of percent of each other. Although in these comparisons the parameters of semi-analytic models were adjusted to match the ionization fractions of the simulations at the redshifts of interest.

\begin{figure*}
  \resizebox{0.5\hsize}{!}{\includegraphics{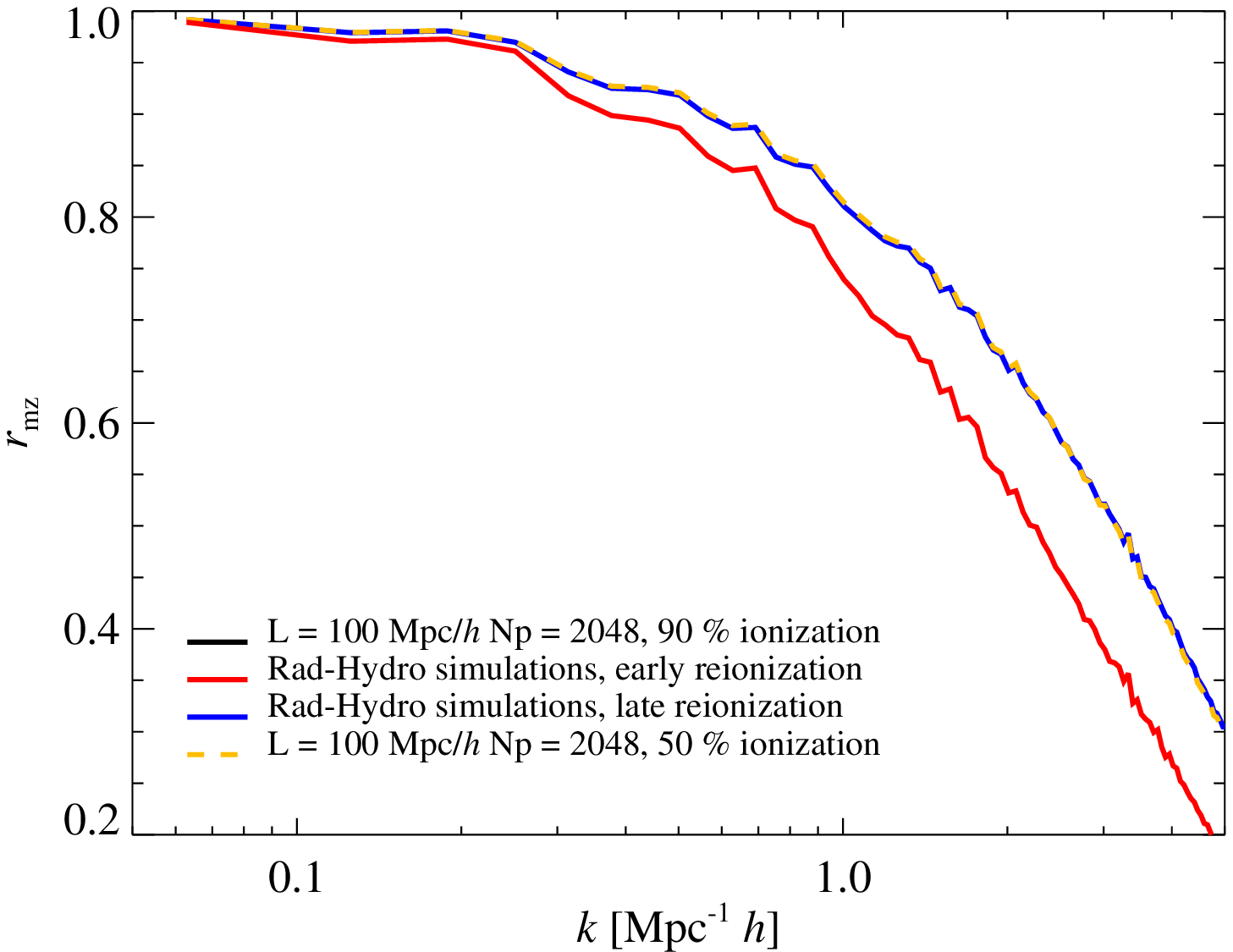}}%
  \resizebox{0.5\hsize}{!}{\includegraphics{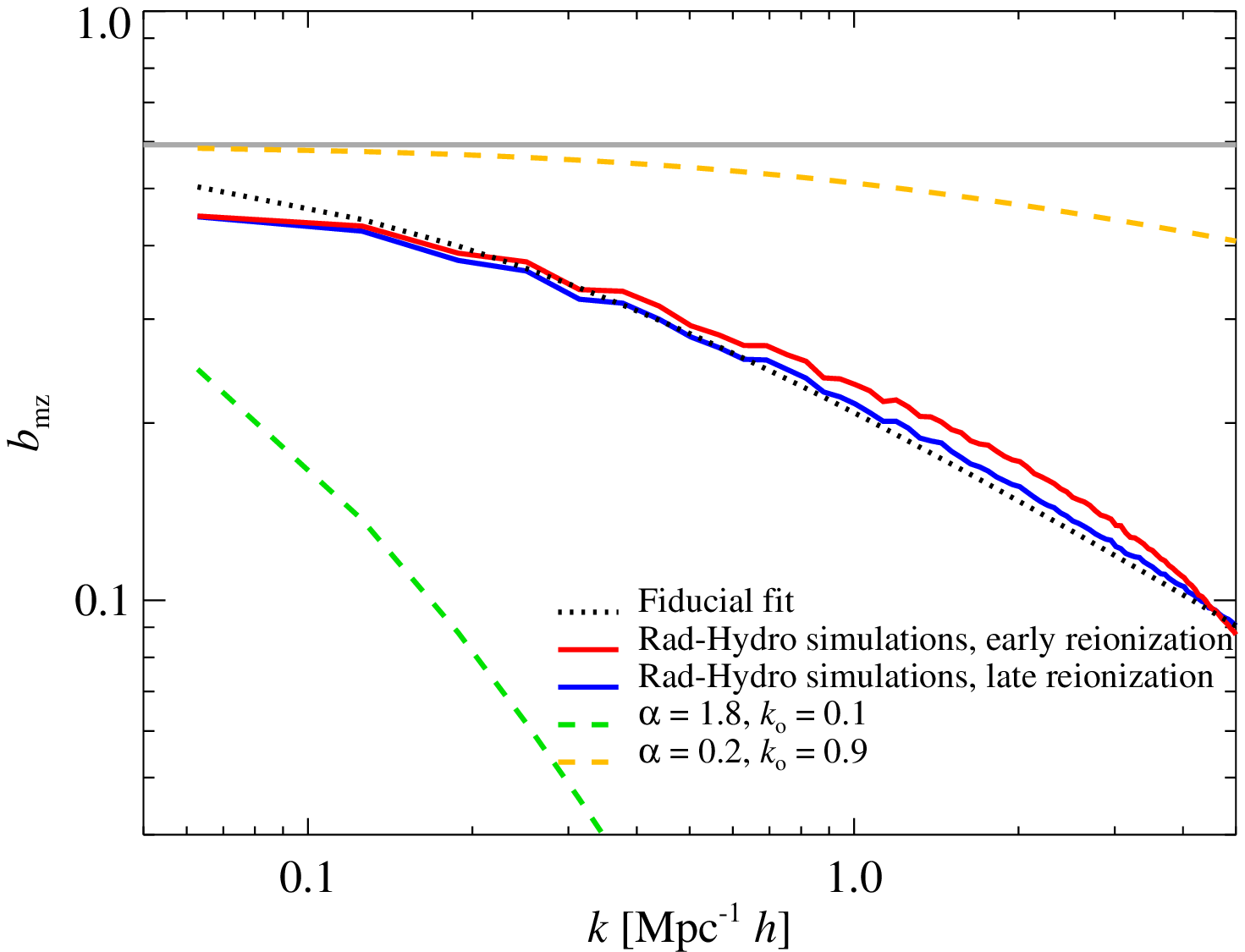}}\\
 \caption{The correlation, $\rmz$, and bias, $\bmz$, calculated from simulations as a function of scale ($k$). Left: Results for $\rmz$ from two different simulated reionization scenarios are shown by the red line(early reionization) and blue line(late reionization) with a resolution of L = 100 Mpc/$h$ box length within 2048 cells (solid lines). In both scenarios $\rmz$ decreases rapidly on scales below 1 Mpc/$h$ and the high resolution simulations show higher correlations to smaller scales. The overall shape of $\rmz$ is largely independent of whether reionization occurred early or late and is completely independent of the ionization threshold chosen (90 - 50\%) for the $\zre$ field (solid blue and dashed orange lines, respectively). Right: Results for $\bmz$ from simulations (solid lines), the fiducial model, which is the best fit $\kb$ and $\al$ parameters to $\bmz$ from simulations (dotted line), and the $\bmz$ functions that represent the long and short duration reionization is the orange and light green dashed line, respectively. The horizontal grey line represents the analytical prediction from spherical collapse.}
\label{fig:correl_bias}
\end{figure*}

In this paper, we present a substantially more accurate semi-analytical model that is statistically informed by simulations with radiative transfer and hydrodynamics. The implementation of this model is fast, versatile and easily applied to large N-body simulations, thus it can be scaled up to the large volumes required by the current and future EoR experiments without loss of accuracy. This paper is the first in a series that explore EoR observables produced via our model. We focus on CMB related observables in \citet{ZR2,ZR3} and the 21cm in \citet{ZR4}. In section~\ref{sec:meth}, we present our fast semi-analytical model and the simulations it is calibrated on. Section~\ref{sec:comp} compares the model to the simulations on a cell by cell bases. We show results on the global reionization history and the typical correlation between ionized regions in section~\ref{sec:res}. We compare to previous work and discuss caveats to our model in section~\ref{sec:dis} and conclude in section~\ref{sec:con}. 
Throughout the paper, we adopt the concordance cosmological parameters: $\Omega_{\rmn{m}} = 0.27$, $\Omega_{\Lambda} = 0.73$, $\Omega_{\rmn{b}} = 0.045$, $h = 0.7$, $n_{\rmn{s}}= 0.96$, and $\sigma_8  = 0.80$. 

\section{Methodology}
\label{sec:meth}

We present a novel semi-analytic model for calculating the evolution of the 3-dimensional ionization field in large volumes $\gtrsim ({\rm Gpc}/h)^3$, which is currently not attainable with direct simulations. Using radiation-hydrodynamic simulations, we demonstrate that the redshift at which a volume-element is ionized can be calculated by filtering a nonlinear density field with a simple parametric function.
Our method can be used to map high-resolution, intermediate-volume radiation-hydrodynamic simulations onto lower-resolution, larger-volume N-body simulations in order to make mock observations and theoretical predictions. In addition, the model parameters can be varied away from the fiducial values in order to explore the reionization parameter space (e.g.~the timing and duration of the EoR).

\subsection{Hydrodynamic Simulations with Radiative Transfer}

We adopt the hybrid approach in simulating cosmic reionization previously described in \citet{Trac2008}. First, a high-resolution N-body simulation is used to evolve the matter distribution and track the formation of dark matter halos. The resulting halo catalogs are used to develop a subgrid model for high-redshift radiation sources. Second, direct RadHydro (radiative transfer + hydrodynamic + N-body) simulations are used to simultaneously solve the coupled evolution of the dark matter, baryons, and radiation. 

A particle-particle-particle-mesh (P$^3$M) code is used to run a high-resolution simulation with $3072^3$ dark matter particles in a $100\ \Mpch$ comoving box. A spherical overdensity halo finder is used on the fly to identify collapsed dark matter halos with average densities equal to 200 times the average cosmic density. With a particle mass resolution of $2.58 \times 10^6\ \Msunh$, we can reliably locate dark matter halos down to the atomic cooling limit ($T \sim 10^4$ K, $M \sim 10^8\ \Msunh$). The halos are populated with radiation sources and the ionizing photon production rate $\dot{N_\gamma}(M,z)$ is calculated using the halo model described in \citet{Trac2007}.

\begin{figure*}
 \begin{minipage}[t]{0.5\textwidth}
   \centering{\it \large Simulation}
 \end{minipage}
 \hfill
 \begin{minipage}[t]{0.5\textwidth}
   \centering{\it \large Model}
 \end{minipage}
  \resizebox{0.5\hsize}{!}{\includegraphics{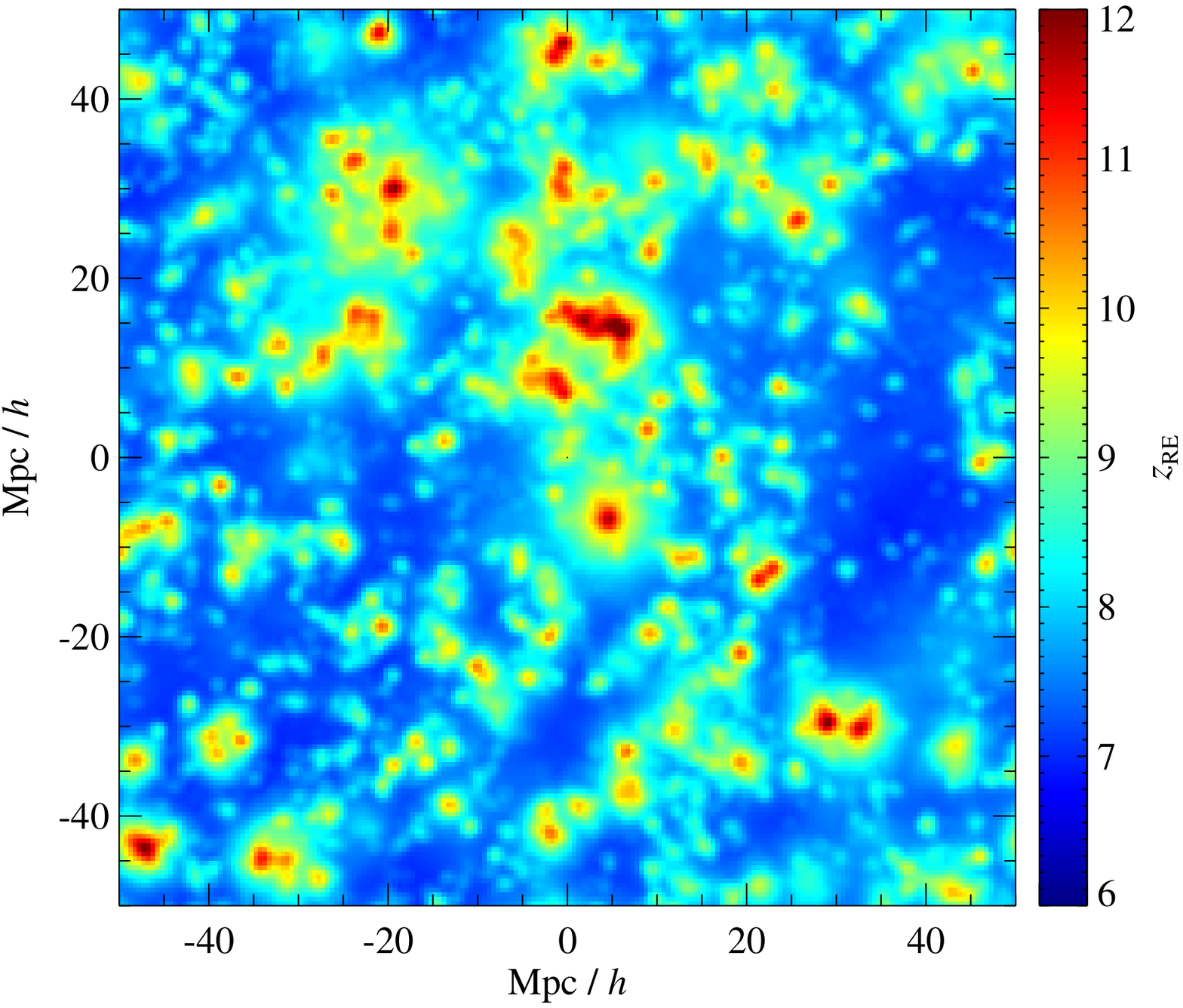}}%
  \resizebox{0.5\hsize}{!}{\includegraphics{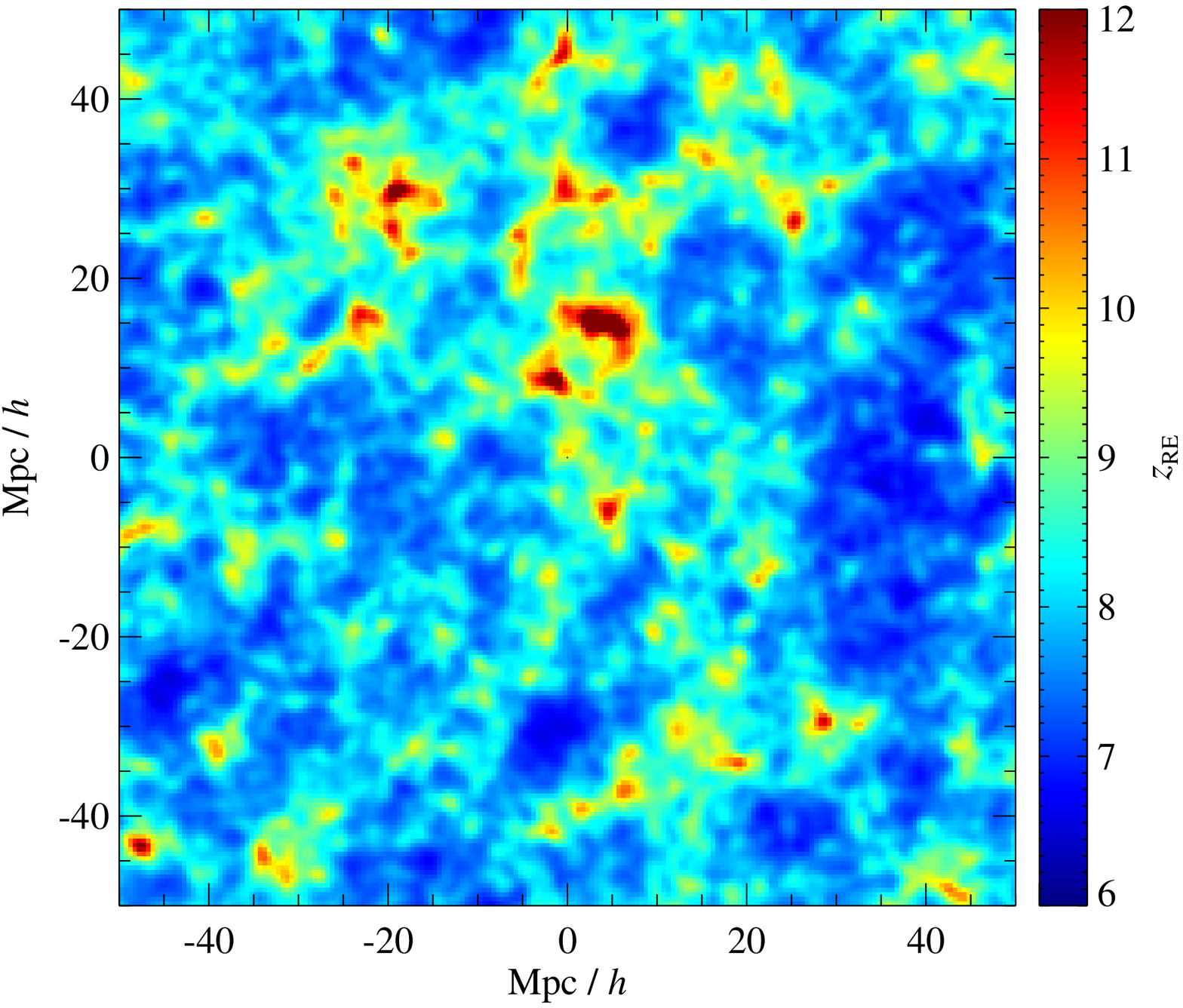}}\\
  \resizebox{0.5\hsize}{!}{\includegraphics{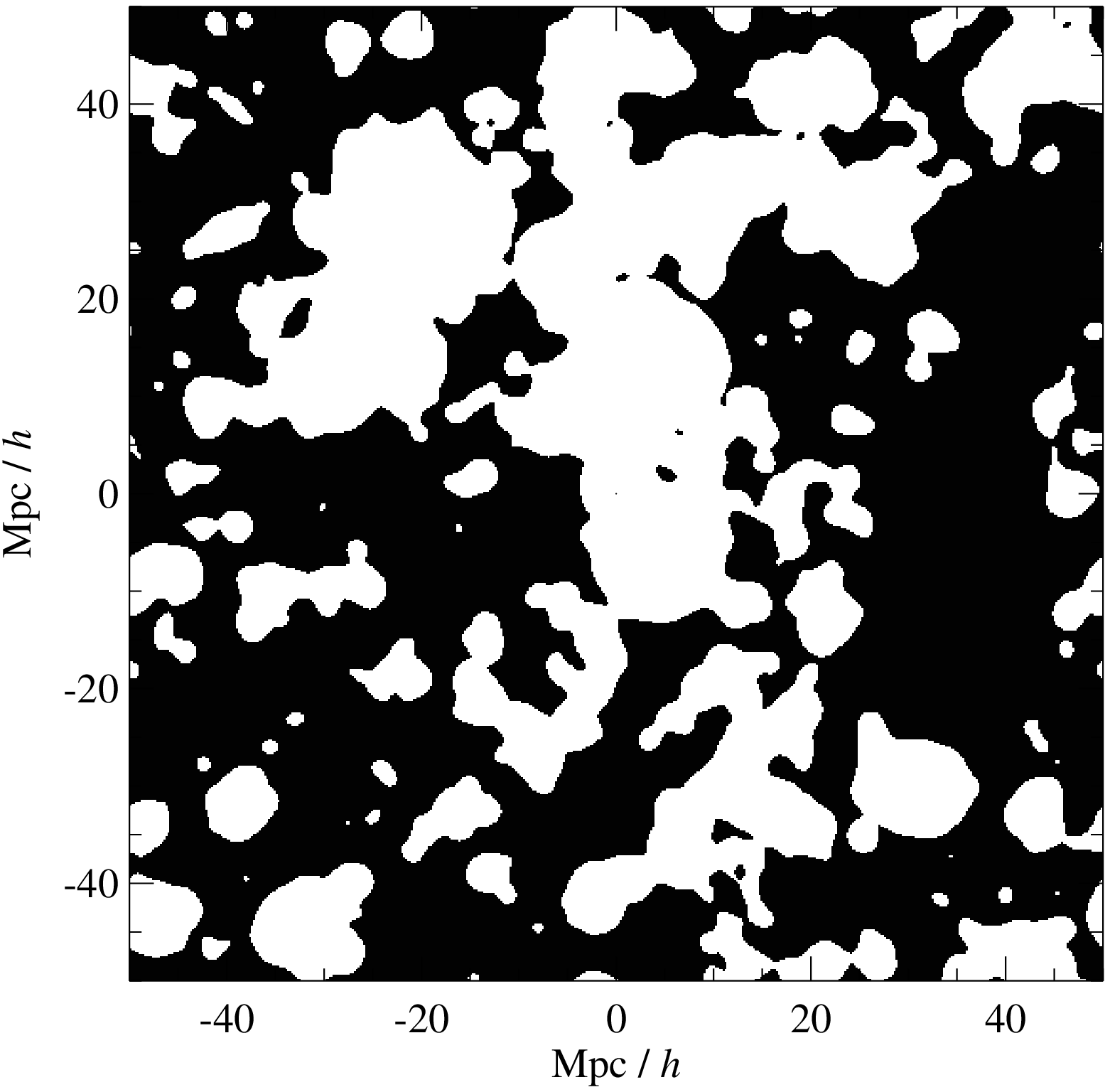}}%
  \resizebox{0.5\hsize}{!}{\includegraphics{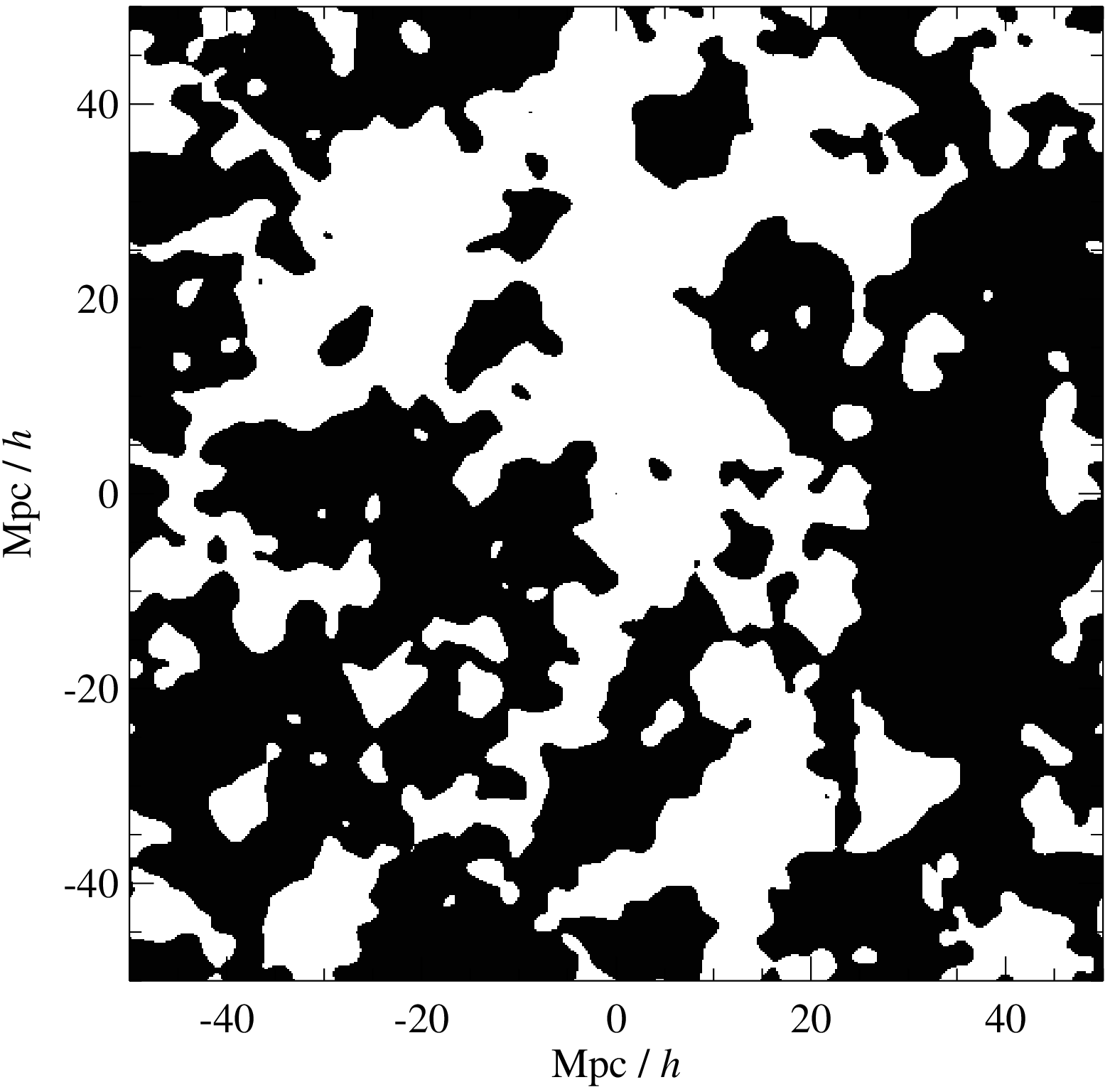}}\\
  \caption{A comparison of  the $\zre$ fields (top panels) and the ionization fraction field at $z=8.1$ (bottom panels) and between our RadHydro simulation with a late reionization scenario (left panels) and model (right panels). The color scales in the top panels illustrate the redshift at which these cells reionize and the black regions in the lower panels correspond to non-ionized regions, while the white regions are ionized $z=8.1$. The simulations were degraded down to a resolution of 1 Mpc/$h$ (see Sec.~\ref{sec:comp} for averaging details) and the slices thicknesses are1 Mpc/$h$. In both upper and lower panels the model captures the same structures as the RadHydro simulation on large scales. The $\zre$ fields differ on small scales in the void and filament regions, where the our method tends to predict earlier redshifts of reionization.} 
\label{fig:sim_comp}
\end{figure*}

The RadHydro code combines a cosmological hydrodynamic code \citep[moving frame hydrodynamics + particle-mesh N-body;][]{Trac2004} with an adaptive raytracing radiative transfer algorithm \citep{Trac2007}. The raytracing algorithm has adaptive splitting and merging and utilizes a two-level radiative transfer (RT) grid scheme to obtain better resolution and scaling. We have run two moderate-resolution RadHydro simulations each with $2048^3$ dark matter particles, $2048^3$ gas cells, and 17 billion adaptive rays. The RadHydro and N-body simulations were run using the Blacklight supercomputer at the Pittsburgh Supercomputing Center (PSC).

In the first RadHydro simulation, reionization occurs earlier with a midpoint of $z \approx 10$ (mass and volume-weighted ionization fractions are $\approx0.5$) and is effectively completed by $z \approx 8.7$ (radiation filling factor of the radiation-hydrodynamic grid reaches unity). This early reionization model has a Thomson optical depth for electron scattering $\te \approx 0.088$, which is in good agreement with current observational constraints. From the WMAP 7-year results, the Thomson optical depth is $\te = 0.088 \pm 0.015$ assuming instantaneous reionization and $\te = 0.087 \pm 0.015$ if the width of reionization is allowed to vary \citep{WMAP7pars}. In the second simulation, reionization occurs later with a midpoint of $z=8$ and finished by $z \approx 6.9$. The late model has $\taue\approx 0.067$, which is lower but within $2\sigma$ of the WMAP best-fit value.

In the two basic simulations, the radiative transfer of the ionizing photons proceeded such that large-scale, overdense regions near sources are generally ionized earlier than large-scale, underdense regions far from sources. The subgrid model for sources only included high-redshift galaxies with Population II stars \citep{Schaerer2003} since they are expected to provide the dominant contribution to the ionizing photon budget. The simulations do not include reionization by Population III stars or X-ray sources, which primarily affect the earliest phase of the EoR \citep[e.g.][]{Furl2006}. We also neglect additional clumping and self-shielding of small-scale dense absorbers such as mini-halos \citep[e.g.][]{Shapiro2004} or Lyman limit systems \citep[e.g.][]{Gnedin2006}. We will explore other reionization scenarios using different models for sources and sinks in future work.

\subsection{Semi-analytic model}

For every cell in the RadHydro simulations, we record the redshift at which it ionizes and construct a reionization-redshift field, $\zre$. Figure \ref{fig:rho_z_sim} shows a slice of the the density and the reionization-redshift fields for the late reionization scenario. For the purposes of this work, regions that are greater than 90\% ionized are considered to be finished reionizing, but we obtain nearly identical statistical quantities when we chose 50\% (cf. Fig~\ref{fig:correl_bias}). Defining a threshold for when cells are considered to be ionized is arbitrary, since there is a sharp transition between the onset of ionization and completion within a cell. Most cells approach 100\% ionization but never reach it due to recombinations. Our results are consistent with \citet{Trac2007} who used a 99\% ionization threshold.

\begin{figure}
\epsscale{1.2}
\plotone{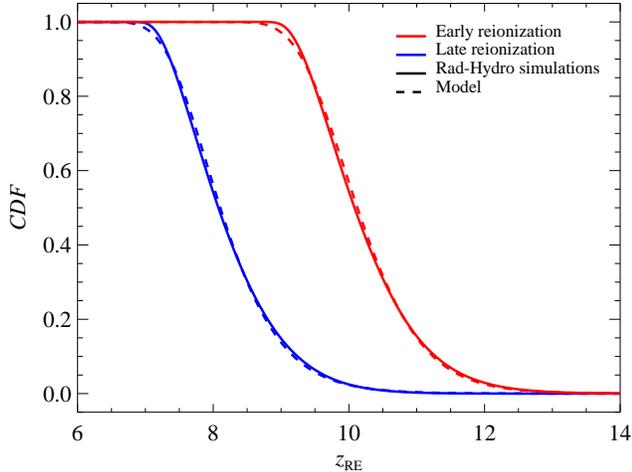}
\caption{A comparison of the cumulative distribution function (CDF) for the $\zre$ field between RadHydro simulations (early and late reionization scenarios, represented by the red and blue lines, respectively) and our model predictions constructed from the same density fields as the simulations. Our models deviate from the RadHydro simulations at the beginning and end of reionization, since the models are constructed for densities fields at $\zbar$. Overall this accounts for percent level changes in the duration of reionization, the exact amount will depend on the definition for the duration.}
\label{fig:comp_zre}
\end{figure}

Figure~\ref{fig:rho_z_sim} shows the over-density field at $z =8$ and the corresponding reionization field from the simulations, which clearly illustrates that the reionization-redshift is associated with the density. The two fields are highly correlated since large-scale over-densities near sources are generally ionized earlier than large-scale under-dense regions far from sources \citep{Bark2004}. Other semi-analytical approaches implicitly invoke this association when constructing their models \citep[e.g.][]{Furl2004,Zahn2005}, and Figure~\ref{fig:rho_z_sim} illustrates that this assumption is fairly accurate down to Mpc scales. Our method quantifies the correlation between the reionization-redshift and density in our RadHydro simulations and uses this correlation and its bias to construct reionization fields from any density field. First, we define the following fluctuations fields

\begin{equation} 
\delta_{\rmn{m}} (\xvec) \equiv \frac{\rho(\xvec) - \bar{\rho}}{ \bar{\rho}},
\label{delm}
\end{equation}
\noindent and
\begin{equation} 
\delta_{\rmn{z}} (\xvec) \equiv \frac{[1 + \zre] - [1 + \zbar]}{1 + \zbar}
\label{delz}
\end{equation}

\noindent where $\bar{\rho}$ is the mean matter density and $\zbar$ is the mean value for the $\zre$ field.
This construction has the advantage that we removed mean redshift dependences of $ \zre$ and these fields are dimensionless.
Then we quantify their correlation using two point statistics $\left<\delta_i\delta_j\right> = P_{ij} (k)$ in Fourier space (i.e. power spectra and cross spectra) assuming isotropy and we calculate their 
cross correlation 

\begin{figure}
\epsscale{1.2}
\plotone{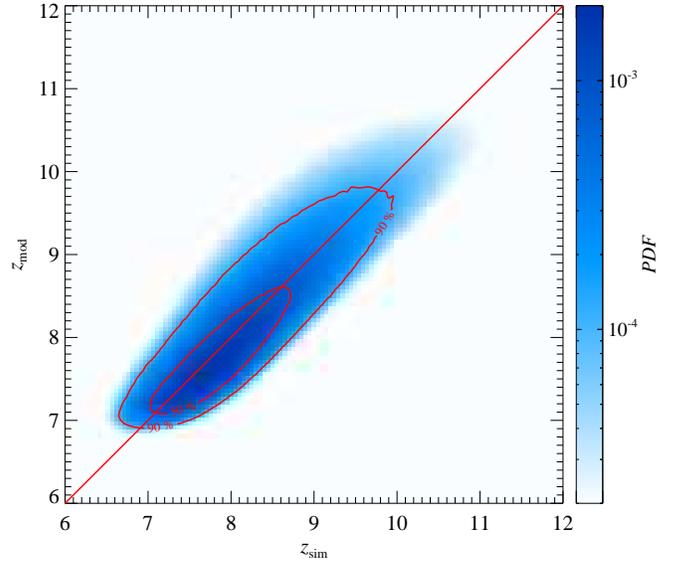}
\caption{The 2D probability distribution function (PDF) between the $\zre$ fields from the RadHydro simulation for late reionization and our model predictions constructed from the same density field compared on the cell by cell level. The contours represent the cumulative probability distribution function for 50\% and 90\% of the cells in this comparison. Our model preforms extremely well around z = $\zbar$ and slightly biases the $\zre$ field to larger redshift values.}
\label{fig:comp_pdf}
\end{figure}

\begin{equation}
\rmz (k) = \frac{P_{\rmn{mz}}(k)} {\sqrt{P_{\rmn{zz}}(k)P_{\rmn{mm}}(k)}}. 
\label{eq:xcor}
\end{equation}

\noindent and linear bias,

\begin{equation}
\bmz = \sqrt{\frac{P_{\rmn{zz}}(k)}{P_{\rmn{mm}}(k)}}.
\label{eq:bias}
\end{equation}

\noindent Figure \ref{fig:correl_bias} show that these fields are highly correlated on scales above 1 Mpc$/h$, and on scales below 1 Mpc$/h$ the correlation decreases. Similar cross correlations between ionization fields and density are calculated in \citet{Zahn2011} for fixed redshifts. As a result of $\rmz (k)$ being highly correlated  on scales above 1 Mpc$/h$, just knowing the bias between these two fields allows one to construct either field from the other by filtering the initial field with the bias. The newly constructed field will statistically match the results from the simulations. We chose to calculate a the linear bias (cf. Eq.~\ref{eq:bias}), since this bias does well to match the $\zre$ from the simulations on scales $>$1 Mpc/$h$ (cf. Fig.~\ref{fig:sim_comp}). Going to higher order correlations and bias models should improve this comparison, but is not necessary to capture the global features at the accuracy we are interested in; such extensions are left for future work. We find that the linear bias can be fit by a three parameter function (cf. Fig.~\ref{fig:correl_bias}). The simple functional form for the bias factor is,

\begin{equation}
\bmz = \frac{\bo}{\left(1 + k/\kb\right)^{\al}} 
\label{eq:bias_form}
\end{equation}

\noindent with the three free parameters being, $\bo$, $\kb$, and $\al$. A simple description for the parameters is as follows:
$\bo$ is the bias amplitude, $\kb$ is the scale threshold, and $\al$ is the asymptotic exponent.
We fit $\kb$, and $\al$ to results from the simulations using least squares fitting with the correlation function ($\rmz$) as weight (cf. Fig.~\ref{fig:correl_bias}). This weighting emphasizes the scales where $\delta_{\rmn{z}}$ and $\delta_{\rmn{m}}$ are highly correlated and down weights the small scales. The best fit values for the $\bmz$ are $\kb = 0.185$ Mpc/$h$ and $\al = 0.564$. Hereafter, we refer to these parameter values as the fiducial bias parameters. We show in Figure~\ref{fig:correl_bias} that our parametric function with these fiducial parameters is comparable to calculated simulation bias. Also we explore the parameter space of $\kb$ and $\al$ and allow them to vary about the fiducial parameters.

The third parameter, $\bo$, which is the bias amplitude on the largest scales is not fit for, since the simulations that we ran have finite box sizes and these scales extend beyond their box size. Figure~\ref{fig:correl_bias} shows that $\bmz$ is asymptotically approaching some value for $\bo$, but fitting for this value would be inaccurate and degenerate with $\kb$ and $\al$.  Instead we refer to previous analytic work by \citet{Bark2004} where this large-scale bias is derived using the excursion set formalism \citep[i.e. extended press-schechter formalism; ][]{Bond1991}.
They show that on these large-scales the differences in the redshift of collapse and reioinization for various over-densites are related via,

\begin{equation}
\delta_{\rmn{z}}(k\rightarrow0) = \frac{\delta_{\rmn{m}}(k\rightarrow0)} { \delta_{\rmn{c}}}.
\end{equation}
\noindent Here $\delta_{\rmn{c}} = 1.68$ is the critical over-density threshold. Throughout this work we use the value $1/ \delta_{\rmn{c}}$ for $\bo = 0.593$.

\subsection{Implementation}

We used a particle-particle-particle-mesh (P$^3$M) N-body code to evolve 2048$^3$ dark matter particles in a 2 Gpc/$h$ box to generate the over-density fields, $\delta_{\rmn{m}}$, down to $z= 5.5$. We convolve the $\delta_{\rmn{m}}$ with a filter consisting three elements: (1) A cubical top hat filter, $\Xi(k)$, which deconvolves the smoothing used to construct $\delta_{\rmn{m}}$ from simulation, (2) A Fourier transform of a real space top hat filter $\Theta(k)$, which smoothes $\delta_{\rmn{m}}$ to resolution of 1 Mpc/$h$, and (3) The bias function from Equation \ref{eq:bias_form}. The assembled filter takes this form 

\begin{equation}
W_{\rmn{z}}(k) \equiv \frac{\bmz \Theta(k)}{\Xi(k)}, 
\label{eq:filter}
\end{equation}

\noindent and we apply this filter at $\zbar$. We Fourier transform back to real space and convert the newly constructed $\delta_{\rmn{z}}$ field to $\zre$ field using Eq.~\ref{delz} with the same $\zbar$ as the density field, here $\zbar$ essentially sets the midpoint of reionization. Thus, we have a complete ionization history for the density field used and from $\zre$ we can construct maps or 3D realization at a given redshift. The entire process is quick, since it only requires Fourier transforms, which can be done with any fast Fourier transform software and scale like NLog(N), where N is the number of cells. We found that I/O of these simulation data files dominates the total time when producing a $\zre$ field. This allows one to easily explore the parameter space of $\kb$, $\al$, and $\zbar$.

\section{Comparison to Simulations}
\label{sec:comp}

\begin{figure}
\epsscale{1.2}
\plotone{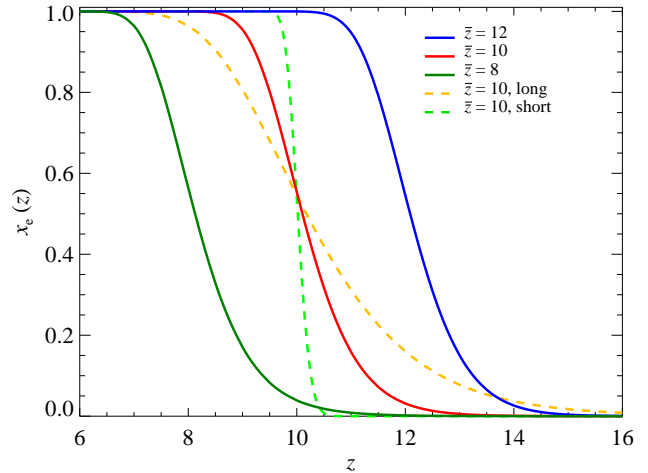}
\caption{The ionization fraction as a function of redshift, $\xe(z)$, for three fiducial parameter models with $\zbar = 8,10,12$ (green,red, and blue lines, respectively) and the long and short duration reionization models (orange and light green dashed line, respectively). The fiducial models have similar ionization histories they are just shifted according to $\zbar$.}
\label{fig:xe}
\end{figure}

We compare the results from the RadHydro simulations for the reionization-redshift fields, $\zre$, to our model predictions that were constructed field from the same simulated density fields. Similar to \citet{Zahn2011}, we also compare their ionization fields at fixed redshifts. Unlike previous comparisons between simulations and semi-analytic models, our comparisons of the $\zre$ fields are computed at the cell by cell level and our model is not tuned to match a particular ionization fraction at a given redshift. Our comparisons go beyond bubble size and distribution tests and looks at the redshift evolution of these quantities as well. In order to make these direct comparisons, the simulations must be smoothed to the scales on which we apply our model $\gtrsim 1$Mpc/$h$, since they are at a higher resolution than 1 Mpc/$h$. The simulations are smoothed by convolving them with a cubical tophat filter, which degrades the resolution of their density and $\zre$ fields to $\sim 1$ Mpc/$h$.  After smoothing the simulations we find that the same structures in the $\zre$ field slice seen in Figure~\ref{fig:rho_z_sim} are still visible in Figure~\ref{fig:sim_comp}.

\begin{figure*}
  \resizebox{0.5\hsize}{!}{\includegraphics{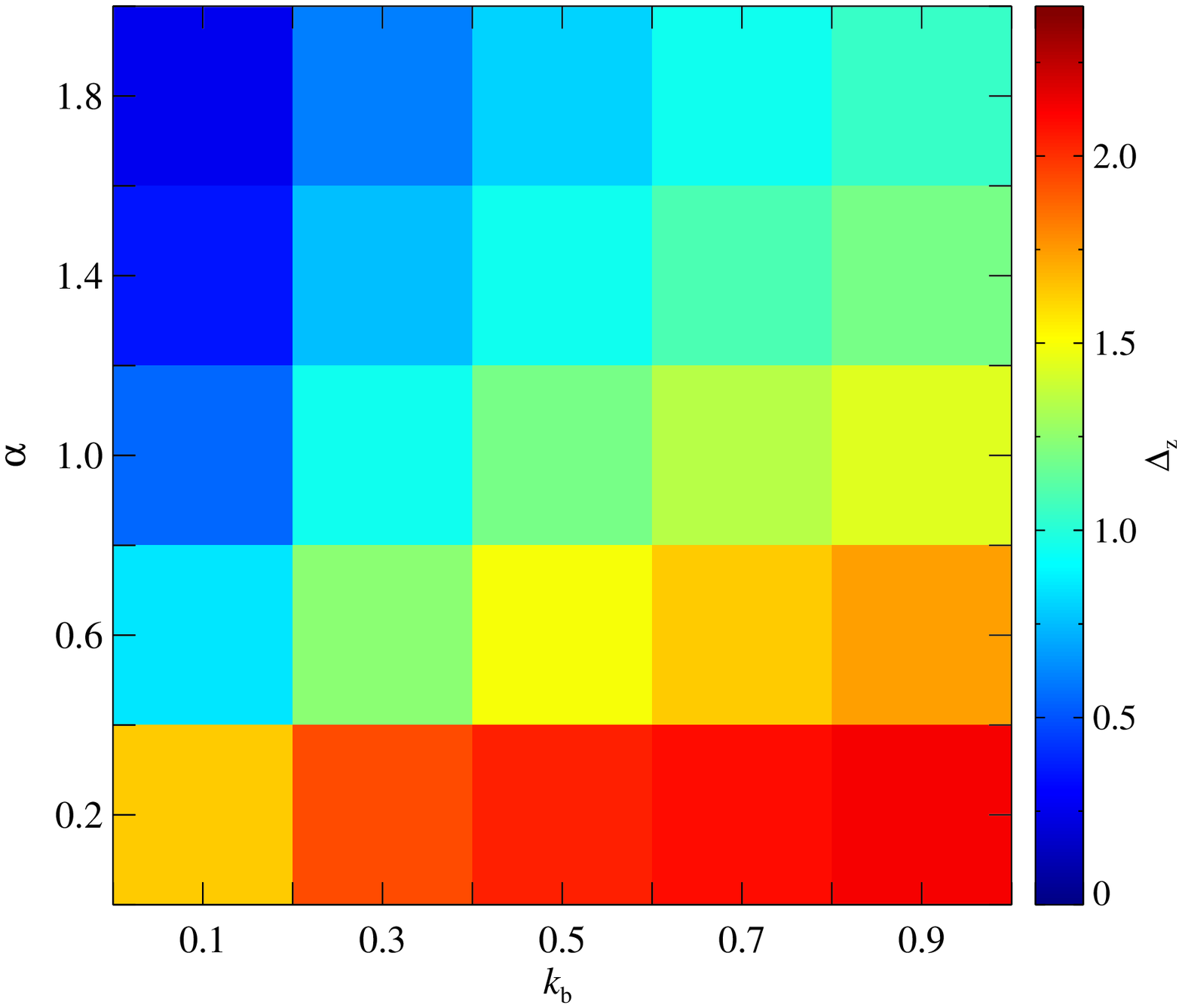}}%
  \resizebox{0.5\hsize}{!}{\includegraphics{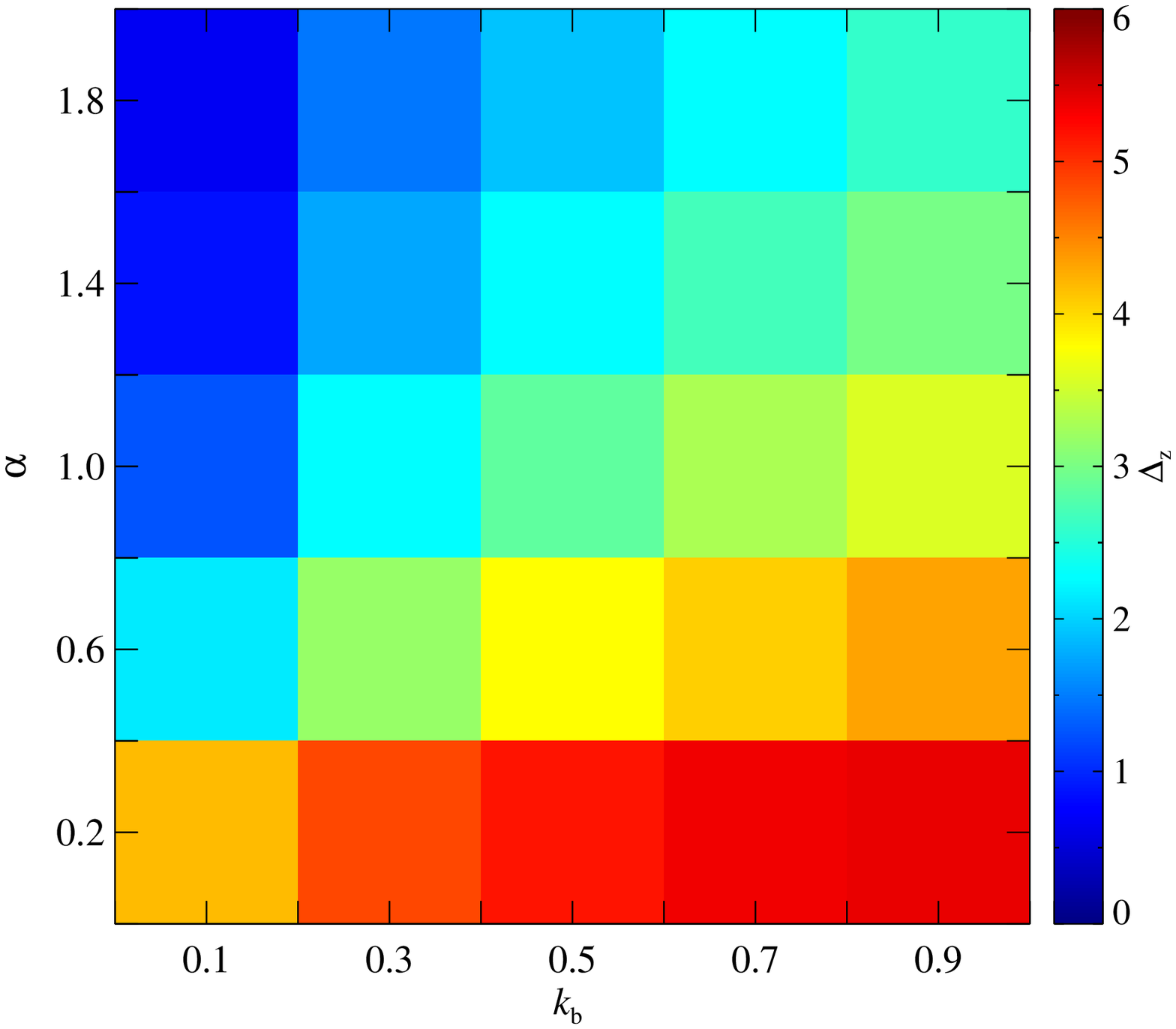}}\\
  \caption{Parameter grids of $\kb$ and $\alpha$ for the width of reionization, $\Delz$. In the left panel $\Delz \equiv z(\xe = 25\%) - z(\xe = 75\%)$ and in the right panel $\Delz \equiv z(\xe = 5\%) - z(\xe = 95\%)$. Increasing $\kb$ extends the duration of reionization, while increasing $\al$ shortens the duration of reionization. The trends for $\Delz (\kb,\al)$ are independent of the definition of $\Delz$. There are degenerate combinations of $\kb$ and $\al$ that give the same values for $\Delz$.}  
\label{fig:xe_grid}
\end{figure*}

Our model predictions for $\zre$ are constructed on the density fields that are smoothed in the same way as the $\zre$ fields, but the $\bmz$ used is calculated directly from the original simulation. The top panels of Figure~\ref{fig:sim_comp} show that the $\zre$ fields from the simulations and our model agree on large scales in their spatial distribution, structure and evolution. Additionally, this illustrates that on smaller scales there are slight disagreements in the filament and void regions as well as the shape of the structures produced from the models, which are more filamentary than simulations. We find that the ionization fields model at $z=$8.1, which approximately corresponds to 50\% ionization in the $\zre$ field, reproduces the simulations results at the same redshift (cf. Fig.~\ref{fig:sim_comp}). The difference found between them are attributed to slightly different mean ionization fractions at $z=$8.1 and the steep changes in these fractions around this redshift. All the differences listed are attributed to the fact that the correlation is not exactly one at these scales, so using a model with a scale-dependent linear bias will not be perfect.

For a quantitative comparison of our model we first compare simulation and model results for the cumulative distribution functions ($CDF$) of $\zre$ for an early and late scenario of reionization. Figure~\ref{fig:comp_zre} shows that our model traces the simulations results well and only deviates from simulations at the very beginning and end of reionization, although this deviation is small. This corresponds to our model having inaccuracies in the densest regions and in the voids, which is illustrated previously in Figure~\ref{fig:sim_comp}. 
The $\zre$ values where the $CDF(\zbar) = 0.5$ in our models show small offsets compared to the simulations $\zre$ values for the same point in the $CDF$. We preform a cell by cell comparison of the $\zre$ fields between simulations and our model predictions, which are the most stringent test. In Figure ~\ref{fig:comp_pdf} we show the 2D probability distribution function (by the blue color scale) and cumulative distribution function (red contours) for the late reionization scenario. A majority of the cells fall along the one to one line in this comparison with some scatter and they are clustered around $\zbar$ of this simulation. We find that 90\% of the cells from the model are within $\sim$ 10\% of the simulation value for their $\zre$ values. After preforming these stringent cell by cell test we did not find any large systematic biases in the mean values or the variations, but these small differences mentioned above are negligible compared to the uncertainties in physics of reionization. We emphasize that no two radiative transfer hydrodynamic simulations nor semi-analytical models have ever been compared by such stringent tests. Thus, our method of filtering the density field with a scale-dependent linear bias is a sufficient description of reionization especially on scales larger than 1 Mpc/$h$.

\section{Results}
\label{sec:res}

In this section, we present results on two physical aspects of the model, the ionization history and the correlation length between ionized regions, which is a proxy for bubble size. We explore the parameter space of Eq.~\ref{eq:bias_form} for aspects, which provides physical understanding of $\zre$ field in relation to the parameters $\kb$, $\al$, and $\zbar$.

\begin{figure*}
  \resizebox{0.5\hsize}{!}{\includegraphics{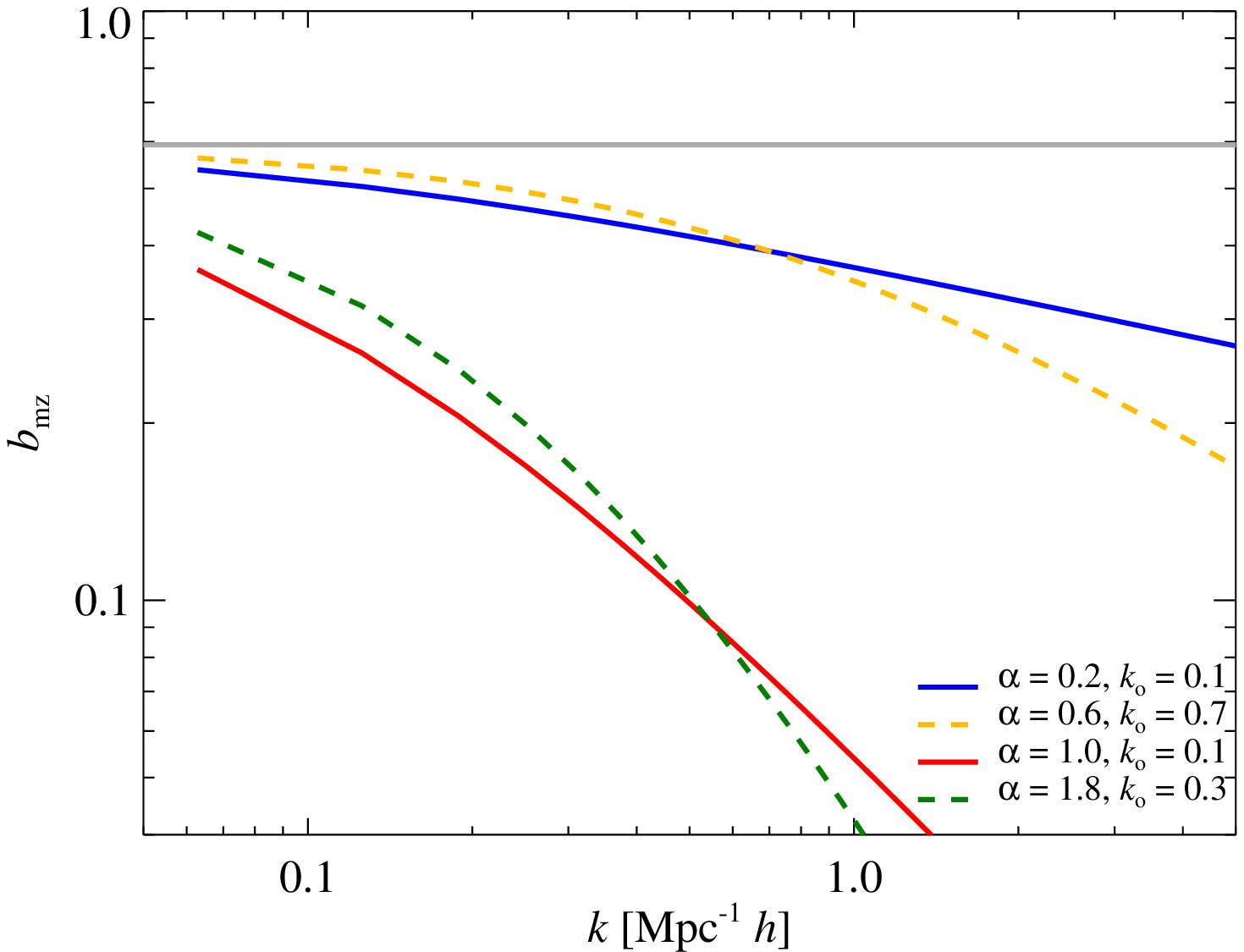}}%
  \resizebox{0.5\hsize}{!}{\includegraphics{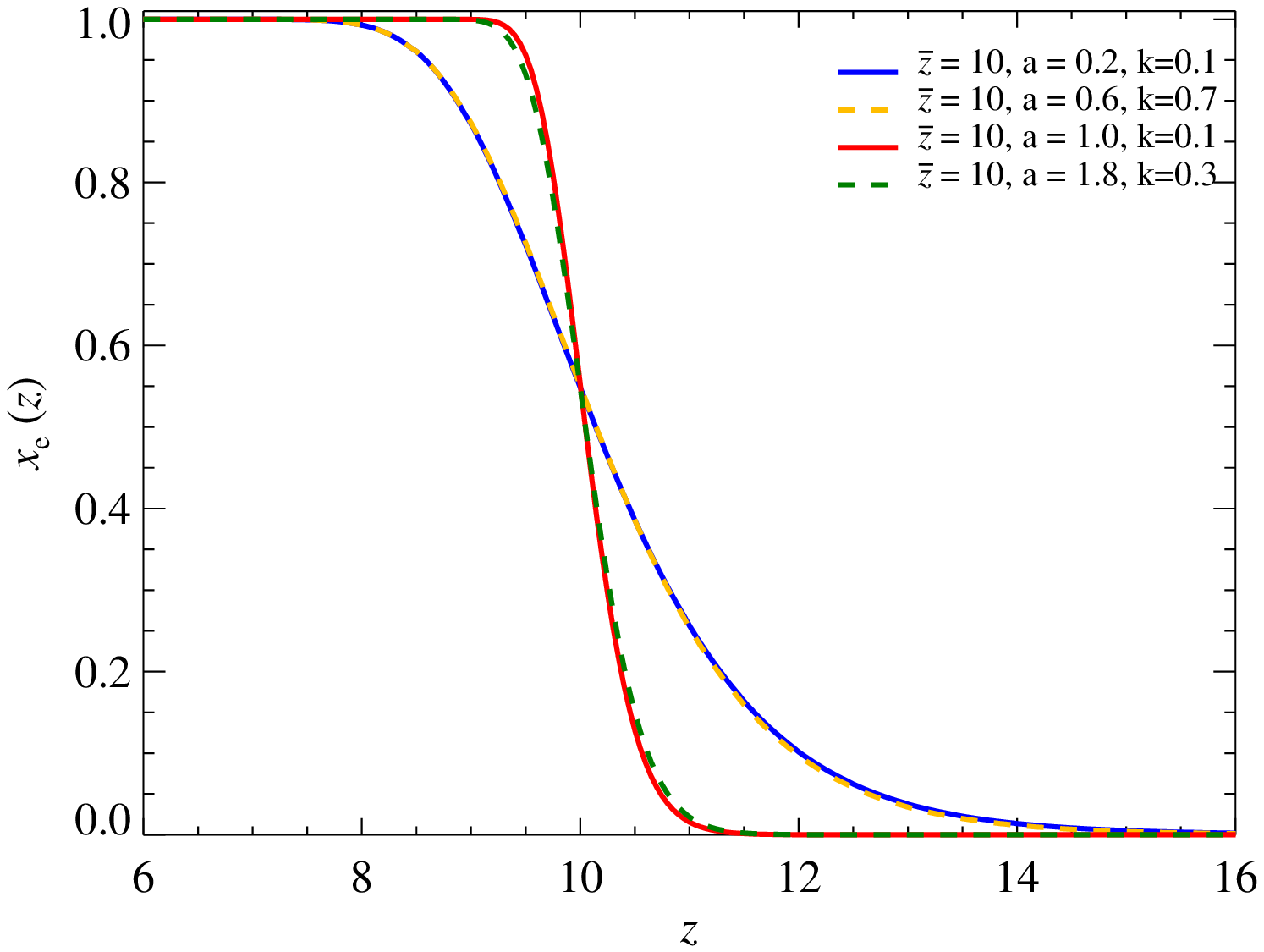}}\\
\caption{Left: The $\bmz$ functions for two examples of degenerate pair combinations of $\kb$ and $\al$ with similar $\Delz$ at fixed $\zbar = 10$ (cf. Fig~\ref{fig:xe_grid}). Right: The 
$\xe(z)$ corresponding to the same pair combinations. Even though the $\bmz$ functions of these degenerate pair combinations are different, they have similar $\xe(z)$.}
\label{fig:xe_comp}
\end{figure*}

\subsection{Ionization History}
\label{sec:delz}
We show the results for the mass weighted ionization history, $\xe$, from our semi-analytic model for various parameter values of $\kb$, $\al$, and $\zbar$ in Fig.~\ref{fig:xe}. In this paper we present only the results for the mass weighted $\xe$, since both the mass and volume weighted $\xe$ have comparable redshift evolution with the slight difference being that the mass weighted $\xe$ increases faster at higher redshifts than the volume weighted $\xe$. The parameter $\zbar$ approximately sets the redshift where $\xe = 50$\%. We find that for a fixed $\kb$ and $\al$ the shape of $\xe(z)$ about $\zbar$ is essentially independent of the value $\zbar$ and the percent differences between $\zbar = 8,10,12$ are below 10\% for $\xe \gtrsim 10$\% (cf. Fig~\ref{fig:xe}). 
The redshift evolution of the ionization history is set by $\kb$ and $\al$, where increasing $\al$ shortens the duration of reionization and increasing $\kb$ lengthens the duration of reionization (cf. Fig.~\ref{fig:xe_grid}). When we increase $\al$, we extend $b_{\rmn{zm}}$ to small scales. Thus, we increase the variance in $\delta_{\rmn{z}}$ and the duration of reionization, while decreasing $\kb$ has opposite effect. These effects can magnified by increasing one parameter while decreasing the other or vice versa.

Similar to previous work \citep[e.g.][]{Zahn2012}, we define two measures for the duration of reionization, $\Delz \equiv z(\xe = 25\%) - z(\xe = 75\%)$ and $\Delz \equiv z(\xe = 5\%) - z(\xe = 95\%)$. In general, semi-analytic models and simulations exclude the early and late times of reionization when defining $\Delz$, since it is difficult to capture the small scale physical processes at these times. Figure~\ref{fig:xe_grid} shows how $\kb$ and $\al$ affect $\Delz$. The trends in $\xe$ from varying the values $\kb$ and $\al$ are the same for $\Delz$ and independent of the definition for $\Delz$. Currently there are upper limits for $\Delz$ CMB small scale measurements \citep{Zahn2012,Mess2012}, independent of definition and lower limits for $\Delz$ from global 21 cm observations \citep{Bow2012}. With a few exceptions, all our models fall well within the constrained region of parameter space for $\Delz$ from these observations, although, we emphasize that the conversion from observation to $\Delz$ is model dependent.

There are degeneracies in both $\xe(z)$ and $\Delz$ in our parametric bias model. Figure ~\ref{fig:xe_comp} shows two examples of parameters pairs values of $\al=0.2,\kb=0.1$ and $\al=0.6,\kb = 0.7$ that have different bias functions, but they have similar ionization histories (cf. Fig.~\ref{fig:xe_comp}) and $\Delz$ values (cf. Fig.~\ref{fig:xe_grid}). These degeneracies are a problem if one wants to relate any $\xe(z)$ observable to the underlying parameters of our model. It is possible to use higher order statistics than $\Delz$, i.e. beyond the variance, to differentiate between these degenerate models, but this does not add physical understanding of how $\al$ and $\kb$ affect EoR. Any measures or proxies for the typical sizes of ionizing regions will differentiate between the degenerate pairs of $\al$ and $\kb$, while providing a more physical understanding of the impact these parameters have on EoR.

\subsection{Correlation Length Between Ionized Regions}

We measure the 3D power spectrum of the ionization field for each EoR model. Here the ionization field is calculated by stepping through redshift and querying each cell $\zre \ge z$ (cf. Fig.~\ref{fig:sim_comp} bottom panels). The power spectrum is expected to peak on scales where the ionized regions are the most correlated. We calculate the wavenumber, $k_e$, where the power spectrum peaks, and define the typical correlation length between ionized regions as $\lambe = 2\pi/k_e$.

\begin{figure}
\epsscale{1.2}
\plotone{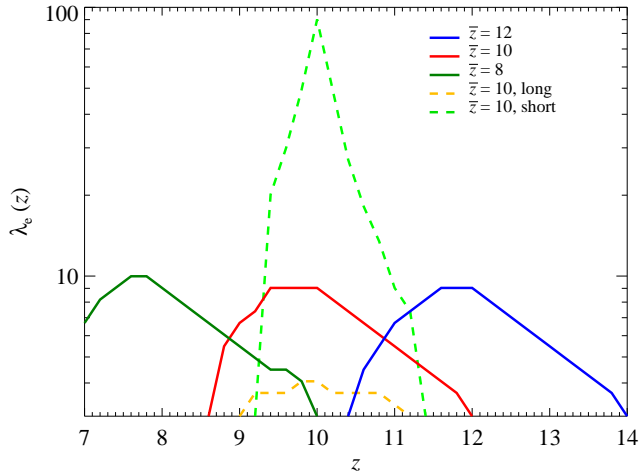}
\caption{The typical correlation length between ionized regions, $\lambe$, for three fiducial parameter models with $\zbar = 8,10,12$ (green,red, and blue lines, respectively) and the long and short duration reionization models (orange and light green dashed line, respectively). The values of $\lambe(z)$ for the fiducial models is independent of $\zbar$. Comparing the values at $z=10$, where all models with $\zbar =10$ have similar $\xe$ values, there is a significant difference between $\lambe(z)$ for the long and short duration reionization.}
\label{fig:bub}
\end{figure}

\begin{figure}
\epsscale{1.2}
\plotone{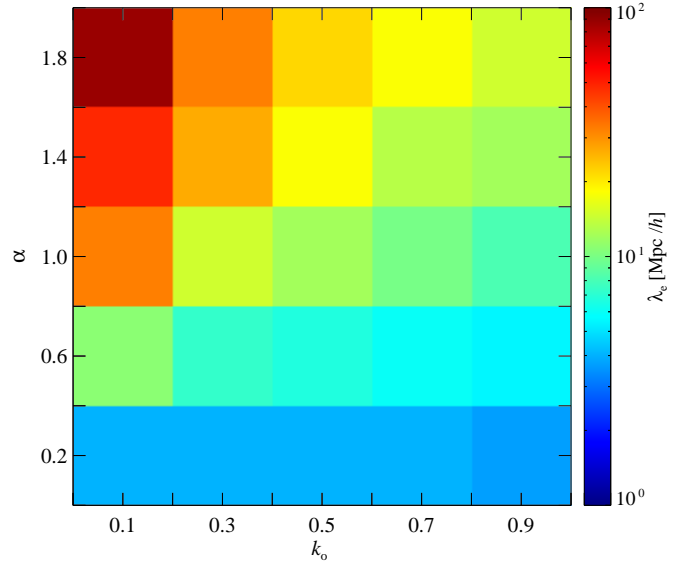}
\caption{Parameter grid of $\kb$ and $\alpha$ for the typical correlation length between ionized regions, $\lambe$ for $\zbar = 10$ and at $z = \zbar$. The previous degenerate combinations of $\kb$ and $\al$ that gave the same values for $\Delz$ give different values for $\lambe$.}
\label{fig:bubgrid}
\end{figure}

The value of $\lambe$ is a proxy for the typical ionized bubble size. We find that the largest $\lambe$ values appear around $\zbar$. After $\zbar$, $\lambe$ no longer measures the typical correlation length between ionized regions, instead it measures the typical correlation between neutral regions (i.e. typical size of neutral clouds).  The redshift evolution of $\lambe$ for fixed $\kb$ and $\al$ is independent of $\zbar$ (cf. Fig.~\ref{fig:bub}). At a fixed $\zbar=10$ the value for $\lambe$ ranges from approximately $3-90$ Mpc/$h$ between our long and short duration reionization models (cf. Fig.~\ref{fig:bub}). Complementary to the results shown in Sec.~\ref{sec:delz}, we find that the long duration reionization model has small correlations lengths between ionized regions, while the short duration reionization model has large correlation lengths between ionized regions.  We chose to compare our models at $\zbar$ when studying the parameter space of $\kb$ and $\al$, since at $\zbar$ all these models have approximately the same ionization fraction ($\sim$~50\%). In Figure~\ref{fig:bubgrid} we show parameter space grid for $\lambe$ a function of $\kb$ and $\al$ for a fixed $\zbar = 10$. The same trends for $\Delz$ as a function of $\kb$ and $\al$ are found in $\lambe$. However, the degenerate combinations of $\kb$ and $\al$ that give similar $\Delz$ values have different $\lambe$ values. Thus, the model degeneracies in any observation that measures $\Delz$ will be removed by any observation that measures $\lambe$ such as the 21cm signal or kSZ power spectrum.

\section{Discusion}
\label{sec:dis}

Semi-analytic models are clearly important tools for understanding the EoR. They are able to quickly explore the large parameter space of unknown and unconstrained physics in an attempt to quantify how these unknown physical processes impact EoR observables. There is an abundance of semi-analytic models \citep[e.g.][]{Furl2004,Zahn2007,Alva2009,Chod2009} that compute an ionization field from a density field (initial conditions or N-body), which rely on a couple of free parameters such as an efficiency $\zeta$ to model the unknown and unresolved physics of EoR. When comparing against simulations, these free parameters are tuned such that the models match the ionization fractions of the radiative transfers simulations. However, one cannot directly compare the parameter values in the model to all values in the simulation and once these parameters are tuned to match simulations their physical interpretations diminish. Our model differs in that all the complex non-linear physics within the simulations is incapsulated in the parametric form $\bmz$. The direct comparison between model and simulations is trivial, since $\bmz$ is computed from the simulations and inserted into our model. In principle, when we calibrated $\bmz$ off simulations there is no loss of accuracy on large scales ($>1$ Mpc/$h$) and there is only one free parameter, $\zbar$. Additionally, our model is extremely quick and easily applicable to large N-body simulations, since it requires only two FFTs to calculate an ionization field, with the rate limiting step being the input and output of the large density and ionization fields, respectively.

The model we proposed is based on an inside-out scenario for reionization, like several other semi-analytic models, which assumes that the first generation of galaxies are the dominant sources for the ionizing photons. This model does not capture exotic reionization scenarios, like those where void regions reionize first. Furthermore, the simulations that our model is derived from do not include ionizing photons from population III stars or first X-ray sources like high redshift quasars  \citep[e.g.][]{Wyit2007,Trac2007}. The impact of these alternate sources for ionizing photons on the EoR is still an open question \citep[e.g.][]{Ahn2012,Visb2012a,Yu2012}. In future work, we look into implementing higher order $\bmz$ and $\rmz$ statistics, which should increase the ability of our model to accurately map simulations onto larger scales.

\section{Conclusions}
\label{sec:con}

We present a novel semi-analytic model for calculating the redshift evolution of the ionization field during EoR that is fast and easily applicable to large volume N-body simulations. Our model is motivated by and calibrated off of RadHydro simulations, which show there is a strong correlation between the density and the reionization-redshift fields on scales $\gtrsim 1$ Mpc/$h$.  A simple filter (cf. Eq.~\ref{eq:filter}) is convolved with a non-linear density field at $z = \zbar$ to obtain a $\zre$, which depends mainly on the parameters $\kb$, $\al$ and $\zbar$. The number of parameters is reduced to one, $\zbar$ (essentially sets the mean reionization-redshift), when the values of $\kb$ and $\al$ are fit to simulation results. 

We found that this model performed well on large scales when we compare it directly to RadHydro simulations. Three of the comparisons we preformed were stringent tests of the simulated and modeled $\zre$ fields: (1) We compared slices of the $\zre$ field between the simulation and model illustrating the minor differences in the evolution of the $\zre$ field. (2) We compared the cumulative distribution of the $\zre$ values, which again yielded minor differences between the simulation and model. (3) We constructed a 2D probability distribution function for a cell by cell comparison of $\zre$ between the simulation and model which showed that 90\% of all $\zre$ values in the model were within 10\% of the simulation values. Given that all the slight difference between the simulation and model were well below the uncertainty in details of the astrophysical processes at work during EoR, the model we proposed is a great tool for incorporating radiation hydrodynamic physics of reionization into large N-body simulations.

We introduce a physical understanding of the parameters $\kb$ and $\al$ by comparing the ionization histories, $\xe$, and typical correlation length between ionized regions, $\lambe$, for various combination of $\kb$ and $\al$. Decreasing the parameter $\kb$ shortens the duration of EoR, while increasing $\kb$ lengthens the duration of EoR, which is a direct result of changing the variance of $\delta_{\rmn{z}}$. The parameter $\al$ has the opposite behavior, i.e. increasing $\al$ shortens the duration of EoR and vise versa. For the values of $\zbar$ we explored, these physical interpretations of $\kb$ and $\al$ are independent of the $\zbar$ value. There are degenerate combinations of $\kb$ and $\al$ that produce nearly identical $\xe(z)$. These degeneracies are broken by $\lambe$ and allow one to further differentiate between parameter combinations. Thus, degeneracy in observables, which depend on $\xe(z)$, can be broken when combined with observables that depend on the $\lambe$. Similar to $\xe(z)$, the values for $\lambe(z)$ at fixed $\kb$ and $\al$ are practically independent of $\zbar$. It is necessary to compare values for $\lambe(z)$ for varying $\kb$ and $\al$ at fixed $\xe$, so comparing at at $z = \zbar$ is a natural choice. For a fixed $z=\zbar$, we demonstrate that smaller $\lambe$ are obtained by increasing $\kb$ or decreasing $\al$ while larger $\lambe$ are obtained by decreasing $\kb$ or increasing $\al$, with values for $\lambe$ ranging from $\sim 3 - 90$ Mpc/$h$. In summary, any combination of $\kb$ and $\al$ that extends the function $\bmz$ to large values of $k$ will increases the amount of small scale structure, thus increasing $\Delz$ and decreasing $\lambe$.

Our method is an accurate and fast tool for exploring galactic reionization of large scales and going forward we will use it to make testable predictions for observables in the CMB and 21cm skies.

\acknowledgments

N.B. is supported by a McWilliams  Center for Cosmology Postdoctoral Fellowship made possible by Bruce and Astrid McWilliams. We thank Aravind Natarajan, Paul La Plante, Jonathan Sievers, and Christian Reichardt,  for useful discussions. We thank N.~Gnedin for his compilation of the ionization and recombination rates and D.~Schaerer for the Population II SEDs. H.T. is supported in part by NSF grant AST-1109730. R.C. is supported in part by NSF grant AST-1108700 and NASA grant NNX12AF91G. A.L. is supported in part by NSF grant AST-0907890 and NASA grants NNX08AL43G and NNA09DB30A. The simulations were performed at the Pittsburgh Supercomputing Center (PSC) and the Princeton Institute for Computational Science and Engineering (PICSciE). We thank Roberto Gomez and Rick Costa at the PSC and Bill Wichser at PICSciE for invaluable help with computing.

\bibliography{nab_PR}

\begin{thebibliography}{64}
\expandafter\ifx\csname natexlab\endcsname\relax\def\natexlab#1{#1}\fi

\bibitem[{{Ahn} {et~al.}(2012){Ahn}, {Iliev}, {Shapiro}, {Mellema}, {Koda}, \&
  {Mao}}]{Ahn2012}
{Ahn}, K., {Iliev}, I.~T., {Shapiro}, P.~R., {Mellema}, G., {Koda}, J., \&
  {Mao}, Y. 2012, \apjl, 756, L16

\bibitem[{{Altay} {et~al.}(2008){Altay}, {Croft}, \& {Pelupessy}}]{Alta2008}
{Altay}, G., {Croft}, R.~A.~C., \& {Pelupessy}, I. 2008, \mnras, 386, 1931

\bibitem[{{Alvarez} {et~al.}(2009){Alvarez}, {Busha}, {Abel}, \&
  {Wechsler}}]{Alva2009}
{Alvarez}, M.~A., {Busha}, M., {Abel}, T., \& {Wechsler}, R.~H. 2009, \apjl,
  703, L167

\bibitem[{{Alvarez} {et~al.}(2006){Alvarez}, {Komatsu}, {Dor{\'e}}, \&
  {Shapiro}}]{Alva2006}
{Alvarez}, M.~A., {Komatsu}, E., {Dor{\'e}}, O., \& {Shapiro}, P.~R. 2006,
  \apj, 647, 840

\bibitem[{{Aubert} \& {Teyssier}(2008)}]{Aub2008}
{Aubert}, D., \& {Teyssier}, R. 2008, \mnras, 387, 295

\bibitem[{{Barkana} \& {Loeb}(2004)}]{Bark2004}
{Barkana}, R., \& {Loeb}, A. 2004, \apj, 609, 474

\bibitem[{{Battaglia} {et~al.}(2012){Battaglia}, {Natarajan}, {Trac}, {Cen}, \&
  {Loeb}}]{ZR3}
{Battaglia}, N., {Natarajan}, A., {Trac}, H., {Cen}, R., \& {Loeb}, A. 2012, in
  prep.

\bibitem[{{Bond} {et~al.}(1991){Bond}, {Cole}, {Efstathiou}, \&
  {Kaiser}}]{Bond1991}
{Bond}, J.~R., {Cole}, S., {Efstathiou}, G., \& {Kaiser}, N. 1991, \apj, 379,
  440

\bibitem[{{Booth} {et~al.}(2009){Booth}, {de Blok}, {Jonas}, \&
  {Fanaroff}}]{Boot2009}
{Booth}, R.~S., {de Blok}, W.~J.~G., {Jonas}, J.~L., \& {Fanaroff}, B. 2009,
  ArXiv e-prints

\bibitem[{{Bowman} {et~al.}(2005){Bowman}, {Morales}, \& {Hewitt}}]{Bow2005}
{Bowman}, J.~D., {Morales}, M.~F., \& {Hewitt}, J.~N. 2005, in Bulletin of the
  American Astronomical Society, Vol.~37, American Astronomical Society Meeting
  Abstracts, 1217

\bibitem[{{Bowman} \& {Rogers}(2012)}]{Bow2012}
{Bowman}, J.~D., \& {Rogers}, A.~E.~E. 2012, ArXiv e-prints

\bibitem[{{Choudhury} {et~al.}(2009){Choudhury}, {Haehnelt}, \&
  {Regan}}]{Chod2009}
{Choudhury}, T.~R., {Haehnelt}, M.~G., \& {Regan}, J. 2009, \mnras, 394, 960

\bibitem[{{Ciardi} {et~al.}(2001){Ciardi}, {Ferrara}, {Marri}, \&
  {Raimondo}}]{Ciar2001}
{Ciardi}, B., {Ferrara}, A., {Marri}, S., \& {Raimondo}, G. 2001, \mnras, 324,
  381

\bibitem[{{Croft} \& {Altay}(2008)}]{Crof2008}
{Croft}, R.~A.~C., \& {Altay}, G. 2008, \mnras, 388, 1501

\bibitem[{{Fan} {et~al.}(2006){Fan}, {Strauss}, {Becker}, {White}, {Gunn},
  {Knapp}, {Richards}, {Schneider}, {Brinkmann}, \& {Fukugita}}]{Fan2006a}
{Fan}, X. {et~al.} 2006, \aj, 132, 117

\bibitem[{{Feng} {et~al.}(2012){Feng}, {Croft}, {Di Matteo}, \&
  {Khandai}}]{Yu2012}
{Feng}, Y., {Croft}, R.~A.~C., {Di Matteo}, T., \& {Khandai}, N. 2012, ArXiv
  e-prints

\bibitem[{{Finlator} {et~al.}(2009){Finlator}, {{\"O}zel}, \&
  {Dav{\'e}}}]{Fin2009}
{Finlator}, K., {{\"O}zel}, F., \& {Dav{\'e}}, R. 2009, \mnras, 393, 1090

\bibitem[{{Furlanetto} {et~al.}(2006){Furlanetto}, {Oh}, \&
  {Briggs}}]{Furl2006}
{Furlanetto}, S.~R., {Oh}, S.~P., \& {Briggs}, F.~H. 2006, \physrep, 433, 181

\bibitem[{{Furlanetto} {et~al.}(2004){Furlanetto}, {Zaldarriaga}, \&
  {Hernquist}}]{Furl2004}
{Furlanetto}, S.~R., {Zaldarriaga}, M., \& {Hernquist}, L. 2004, \apj, 613, 1

\bibitem[{{Geil} \& {Wyithe}(2008)}]{Geil2008}
{Geil}, P.~M., \& {Wyithe}, J.~S.~B. 2008, \mnras, 386, 1683

\bibitem[{{Gnedin} \& {Abel}(2001)}]{Gned2001}
{Gnedin}, N.~Y., \& {Abel}, T. 2001, \na, 6, 437

\bibitem[{{Gnedin} \& {Fan}(2006)}]{Gnedin2006}
{Gnedin}, N.~Y., \& {Fan}, X. 2006, \apj, 648, 1

\bibitem[{{Gruzinov} \& {Hu}(1998)}]{Gruz1998}
{Gruzinov}, A., \& {Hu}, W. 1998, \apj, 508, 435

\bibitem[{{Harker} {et~al.}(2010){Harker}, {Zaroubi}, {Bernardi}, {Brentjens},
  {de Bruyn}, {Ciardi}, {Jeli{\'c}}, {Koopmans}, {Labropoulos}, {Mellema},
  {Offringa}, {Pandey}, {Pawlik}, {Schaye}, {Thomas}, \&
  {Yatawatta}}]{Hark2010}
{Harker}, G. {et~al.} 2010, \mnras, 405, 2492

\bibitem[{{Iliev} {et~al.}(2006){Iliev}, {Mellema}, {Pen}, {Merz}, {Shapiro},
  \& {Alvarez}}]{iliv2006}
{Iliev}, I.~T., {Mellema}, G., {Pen}, U.-L., {Merz}, H., {Shapiro}, P.~R., \&
  {Alvarez}, M.~A. 2006, \mnras, 369, 1625

\bibitem[{{Iliev} {et~al.}(2007){Iliev}, {Pen}, {Bond}, {Mellema}, \&
  {Shapiro}}]{iliv2007}
{Iliev}, I.~T., {Pen}, U.-L., {Bond}, J.~R., {Mellema}, G., \& {Shapiro}, P.~R.
  2007, \apj, 660, 933

\bibitem[{{Knox} {et~al.}(1998){Knox}, {Scoccimarro}, \& {Dodelson}}]{Knox1998}
{Knox}, L., {Scoccimarro}, R., \& {Dodelson}, S. 1998, Physical Review Letters,
  81, 2004

\bibitem[{{La Plante} {et~al.}(2012){La Plante}, {Battaglia}, {Trac}, {Cen}, \&
  {Loeb}}]{ZR4}
{La Plante}, P., {Battaglia}, N., {Trac}, H., {Cen}, R., \& {Loeb}, A. 2012, in
  prep.

\bibitem[{{Larson} {et~al.}(2011){Larson}, {Dunkley}, {Hinshaw}, {Komatsu},
  {Nolta}, {Bennett}, {Gold}, {Halpern}, {Hill}, {Jarosik}, {Kogut}, {Limon},
  {Meyer}, {Odegard}, {Page}, {Smith}, {Spergel}, {Tucker}, {Weiland},
  {Wollack}, \& {Wright}}]{WMAP7pars}
{Larson}, D. {et~al.} 2011, \apjs, 192, 16

\bibitem[{{Lidz} {et~al.}(2006){Lidz}, {Oh}, \& {Furlanetto}}]{Lidz2006}
{Lidz}, A., {Oh}, S.~P., \& {Furlanetto}, S.~R. 2006, \apjl, 639, L47

\bibitem[{{Loeb} \& {Furlanetto}(2013)}]{Loeb2013}
{Loeb}, A., \& {Furlanetto}, S. 2013, Princeton University Press, in press

\bibitem[{{Maselli} {et~al.}(2003){Maselli}, {Ferrara}, \& {Ciardi}}]{Mase2003}
{Maselli}, A., {Ferrara}, A., \& {Ciardi}, B. 2003, \mnras, 345, 379

\bibitem[{{McQuinn} {et~al.}(2005){McQuinn}, {Furlanetto}, {Hernquist}, {Zahn},
  \& {Zaldarriaga}}]{McQn2005}
{McQuinn}, M., {Furlanetto}, S.~R., {Hernquist}, L., {Zahn}, O., \&
  {Zaldarriaga}, M. 2005, \apj, 630, 643

\bibitem[{{McQuinn} {et~al.}(2007){McQuinn}, {Lidz}, {Zahn}, {Dutta},
  {Hernquist}, \& {Zaldarriaga}}]{McQn2007}
{McQuinn}, M., {Lidz}, A., {Zahn}, O., {Dutta}, S., {Hernquist}, L., \&
  {Zaldarriaga}, M. 2007, \mnras, 377, 1043

\bibitem[{{Mellema} {et~al.}(2006){Mellema}, {Iliev}, {Pen}, \&
  {Shapiro}}]{Mell2006}
{Mellema}, G., {Iliev}, I.~T., {Pen}, U.-L., \& {Shapiro}, P.~R. 2006, \mnras,
  372, 679

\bibitem[{{Mellema} {et~al.}(2012){Mellema}, {Koopmans}, {Abdalla}, {Bernardi},
  {Ciardi}, {Daiboo}, {de Bruyn}, {Datta}, {Falcke}, {Ferrara}, {Iliev},
  {Iocco}, {Jeli{\'c}}, {Jensen}, {Joseph}, {Kloeckner}, {Labroupoulos},
  {Meiksin}, {Mesinger}, {Offringa}, {Pandey}, {Pritchard}, {Santos},
  {Schwarz}, {Semelin}, {Vedantham}, {Yatawatta}, \& {Zaroubi}}]{Mell2012}
{Mellema}, G. {et~al.} 2012, ArXiv e-prints

\bibitem[{{Mesinger} \& {Furlanetto}(2007)}]{Mess2007}
{Mesinger}, A., \& {Furlanetto}, S. 2007, \apj, 669, 663

\bibitem[{{Mesinger} {et~al.}(2011){Mesinger}, {Furlanetto}, \&
  {Cen}}]{Mess2011}
{Mesinger}, A., {Furlanetto}, S., \& {Cen}, R. 2011, \mnras, 411, 955

\bibitem[{{Mesinger} {et~al.}(2012){Mesinger}, {McQuinn}, \&
  {Spergel}}]{Mess2012}
{Mesinger}, A., {McQuinn}, M., \& {Spergel}, D.~N. 2012, \mnras, 422, 1403

\bibitem[{{Morales} \& {Wyithe}(2010)}]{Mora2010}
{Morales}, M.~F., \& {Wyithe}, J.~S.~B. 2010, \araa, 48, 127

\bibitem[{{Natarajan} {et~al.}(2012){Natarajan}, {Battaglia}, {Trac}, {Pen}, \&
  {Loeb}}]{ZR2}
{Natarajan}, A., {Battaglia}, N., {Trac}, H., {Pen}, U., \& {Loeb}, A. 2012, in
  prep.

\bibitem[{{Oh} \& {Furlanetto}(2005)}]{Oh2005}
{Oh}, S.~P., \& {Furlanetto}, S.~R. 2005, \apjl, 620, L9

\bibitem[{{Parsons} {et~al.}(2010){Parsons}, {Backer}, {Foster}, {Wright},
  {Bradley}, {Gugliucci}, {Parashare}, {Benoit}, {Aguirre}, {Jacobs},
  {Carilli}, {Herne}, {Lynch}, {Manley}, \& {Werthimer}}]{Pars2010}
{Parsons}, A.~R. {et~al.} 2010, \aj, 139, 1468

\bibitem[{{Pen} {et~al.}(2009){Pen}, {Chang}, {Hirata}, {Peterson}, {Roy},
  {Gupta}, {Odegova}, \& {Sigurdson}}]{Pen2009}
{Pen}, U.-L., {Chang}, T.-C., {Hirata}, C.~M., {Peterson}, J.~B., {Roy}, J.,
  {Gupta}, Y., {Odegova}, J., \& {Sigurdson}, K. 2009, \mnras, 399, 181

\bibitem[{{Petkova} \& {Springel}(2009)}]{Pet2009}
{Petkova}, M., \& {Springel}, V. 2009, \mnras, 396, 1383

\bibitem[{{Santos} {et~al.}(2003){Santos}, {Cooray}, {Haiman}, {Knox}, \&
  {Ma}}]{Sant2003}
{Santos}, M.~G., {Cooray}, A., {Haiman}, Z., {Knox}, L., \& {Ma}, C.-P. 2003,
  \apj, 598, 756

\bibitem[{{Santos} {et~al.}(2010){Santos}, {Ferramacho}, {Silva}, {Amblard}, \&
  {Cooray}}]{Sant2010}
{Santos}, M.~G., {Ferramacho}, L., {Silva}, M.~B., {Amblard}, A., \& {Cooray},
  A. 2010, \mnras, 406, 2421

\bibitem[{{Schaerer}(2003)}]{Schaerer2003}
{Schaerer}, D. 2003, \aap, 397, 527

\bibitem[{{Scott} \& {Rees}(1990)}]{Scot1990}
{Scott}, D., \& {Rees}, M.~J. 1990, \mnras, 247, 510

\bibitem[{{Shapiro} {et~al.}(2004){Shapiro}, {Iliev}, \& {Raga}}]{Shapiro2004}
{Shapiro}, P.~R., {Iliev}, I.~T., \& {Raga}, A.~C. 2004, \mnras, 348, 753

\bibitem[{{Shaver} {et~al.}(1999){Shaver}, {Windhorst}, {Madau}, \& {de
  Bruyn}}]{Shav1999}
{Shaver}, P.~A., {Windhorst}, R.~A., {Madau}, P., \& {de Bruyn}, A.~G. 1999,
  \aap, 345, 380

\bibitem[{{Thomas} {et~al.}(2009){Thomas}, {Zaroubi}, {Ciardi}, {Pawlik},
  {Labropoulos}, {Jeli{\'c}}, {Bernardi}, {Brentjens}, {de Bruyn}, {Harker},
  {Koopmans}, {Mellema}, {Pandey}, {Schaye}, \& {Yatawatta}}]{Thom2009}
{Thomas}, R.~M. {et~al.} 2009, \mnras, 393, 32

\bibitem[{{Trac} \& {Cen}(2007)}]{Trac2007}
{Trac}, H., \& {Cen}, R. 2007, \apj, 671, 1

\bibitem[{{Trac} {et~al.}(2008){Trac}, {Cen}, \& {Loeb}}]{Trac2008}
{Trac}, H., {Cen}, R., \& {Loeb}, A. 2008, \apjl, 689, L81

\bibitem[{{Trac} \& {Pen}(2004)}]{Trac2004}
{Trac}, H., \& {Pen}, U.-L. 2004, \na, 9, 443

\bibitem[{{Trac} \& {Gnedin}(2011)}]{Trac2011}
{Trac}, H.~Y., \& {Gnedin}, N.~Y. 2011, Advanced Science Letters, 4, 228

\bibitem[{{Valageas} {et~al.}(2001){Valageas}, {Balbi}, \& {Silk}}]{Vala2001}
{Valageas}, P., {Balbi}, A., \& {Silk}, J. 2001, \aap, 367, 1

\bibitem[{{Visbal} \& {Loeb}(2012)}]{Visb2012a}
{Visbal}, E., \& {Loeb}, A. 2012, \jcap, 5, 7

\bibitem[{{Wyithe} \& {Cen}(2007)}]{Wyit2007}
{Wyithe}, J.~S.~B., \& {Cen}, R. 2007, \apj, 659, 890

\bibitem[{{Zahn} {et~al.}(2007){Zahn}, {Lidz}, {McQuinn}, {Dutta}, {Hernquist},
  {Zaldarriaga}, \& {Furlanetto}}]{Zahn2007}
{Zahn}, O., {Lidz}, A., {McQuinn}, M., {Dutta}, S., {Hernquist}, L.,
  {Zaldarriaga}, M., \& {Furlanetto}, S.~R. 2007, \apj, 654, 12

\bibitem[{{Zahn} {et~al.}(2011){Zahn}, {Mesinger}, {McQuinn}, {Trac}, {Cen}, \&
  {Hernquist}}]{Zahn2011}
{Zahn}, O., {Mesinger}, A., {McQuinn}, M., {Trac}, H., {Cen}, R., \&
  {Hernquist}, L.~E. 2011, \mnras, 414, 727

\bibitem[{{Zahn} {et~al.}(2012){Zahn}, {Reichardt}, {Shaw}, {Lidz}, {Aird},
  {Benson}, {Bleem}, {Carlstrom}, {Chang}, {Cho}, {Crawford}, {Crites}, {de
  Haan}, {Dobbs}, {Dor{\'e}}, {Dudley}, {George}, {Halverson}, {Holder},
  {Holzapfel}, {Hoover}, {Hou}, {Hrubes}, {Joy}, {Keisler}, {Knox}, {Lee},
  {Leitch}, {Lueker}, {Luong-Van}, {McMahon}, {Mehl}, {Meyer}, {Millea},
  {Mohr}, {Montroy}, {Natoli}, {Padin}, {Plagge}, {Pryke}, {Ruhl}, {Schaffer},
  {Shirokoff}, {Spieler}, {Staniszewski}, {Stark}, {Story}, {van Engelen},
  {Vanderlinde}, {Vieira}, \& {Williamson}}]{Zahn2012}
{Zahn}, O. {et~al.} 2012, \apj, 756, 65

\bibitem[{{Zahn} {et~al.}(2005){Zahn}, {Zaldarriaga}, {Hernquist}, \&
  {McQuinn}}]{Zahn2005}
{Zahn}, O., {Zaldarriaga}, M., {Hernquist}, L., \& {McQuinn}, M. 2005, \apj,
  630, 657

\bibitem[{{Zaldarriaga} {et~al.}(2004){Zaldarriaga}, {Furlanetto}, \&
  {Hernquist}}]{Zald2004}
{Zaldarriaga}, M., {Furlanetto}, S.~R., \& {Hernquist}, L. 2004, \apj, 608, 622

\end{thebibliography}
\bibliographystyle{apj}

\end{document}